\documentclass[prl,aps,twocolumn,superscriptaddress]{revtex4-2}

\usepackage[]{amsmath}
\usepackage[]{braket}
\usepackage[]{amssymb}
\usepackage[]{bm}
\usepackage{graphicx}
\usepackage[colorlinks=true]{hyperref}

\newcommand{\ux}{\bm{u}_x}
\newcommand{\uy}{\bm{u}_y}
\newcommand{\uz}{\bm{u}_z}
\newcommand{\identity}{\mbox{1}\hspace{-0.25em}\mbox{l}}

\usepackage{fontawesome}

\newcommand{\figref}[1]{\figurename~\ref{#1}} 

\begin{document}
\title{Twisting Optomechanical Cavity}
\author{Daigo Oue}
\affiliation{Instituto Superior Técnico, University of Lisbon, 1049-001 Lisboa, Portugal}
\affiliation{Kavli Institute for Theoretical Sciences, University of Chinese Academy of Sciences, Beijing 100190, China}
\affiliation{Department of Physics, Imperial College London, London SW7 2AZ, UK}
\author{Mamoru Matsuo}
\affiliation{Kavli Institute for Theoretical Sciences, University of Chinese Academy of Sciences, Beijing 100190, China}
\affiliation{CAS Center for Excellence in Topological Quantum Computation, University of Chinese Academy of Sciences, Beijing 100190, China}
\affiliation{Advanced Science Research Center, Japan Atomic Energy Agency, Tokai, 319-1195, Japan}
\date{\today}

\begin{abstract}
  Mechanical rotation and oscillation have far lower frequencies than light does;
  thus they are not coupled to each other conventionally.
  In this Letter,
  we show the torsional mechanical oscillation of an optical cavity can be coupled to the optical modes by introducing birefringence, 
  which produces nondegenerate modes in the cavity: ordinary and extraordinary rays.
  Twisting the cavity mixes them and modulates the electromagnetic energy.
  We find torsional optomechanical Hamiltonian by quantising the total energy and reveal the torsional oscillation can be resonantly driven by light.
\end{abstract}

\maketitle

\label{---introduction---}
\par \textit{Introduction.---}%
Radiation force, 
which is discovered by Nichols and Hull~\cite{nichols1903pressure, nichols1903pressure2nd} more than 100 years ago,
is one consequence of the interaction of light with mechanical degrees of freedom.
Momentum per photon is so small (e.g., for visible light, $p = \hbar |\bm{k}| \approx 10^{-27}\ \mathrm{kg\cdot m/s}$),
and the momentum transfer rate from photon to matter is also low in free space.
This is because the frequency of light ($\sim 10^{15}\ \mathrm{Hz}$) is much higher than that of mechanical oscillation ($\sim 10^9\ \mathrm{Hz}$),
and the mode volume of light is quite different from that of mechanical oscillation.
Therefore, large photon flux is necessary to drive a macroscopic object by radiation force,
but high-intensity light source was not available at that time,
and it was difficult even to detect the radiation force.
However, the invention of the laser pioneered the research field of optical trapping and manipulation that micro- or nano- particles can be optically manipulated by utilising the radiation force~\cite{ashkin1970acceleration, ashkin1986observation},
and thus the technique of optical manipulation has been applied not only in physics but also in a wide range of scientific fields such as chemistry and biological science~\cite{gao2017optical}.

Recently, it is realised that photon and quantised mechanical oscillation are coupled to each other by confining both of them in a so-called optomechanical cavity~\cite{aspelmeyer2014cavity}.
The optomechanical cavity can be composed of a pair of mirrors one of which is suspended by a spring and movable.
There are many architectures to confine optical and mechanical oscillation in the same place.
For example,
microdisk resonators~\cite{%
hofer2010cavity,%
ding2011wavelength,%
forstner2012cavity,%
verhagen2012quantum}
and periodic photonic structures~\cite{%
chan2011laser,%
eichenfield2009picogram,%
gavartin2011optomechanical,%
maldovan2006simultaneous,%
safavi2010design,%
eichenfield2009optomechanical}
have also been utilised,
where the elastic vibrations play machanical roles.
Alternatively,
the optical and mechanical coupling have also been demonstrated by using a suspended membranes~\cite{%
thompson2008strong,%
wilson2009cavity,%
favero2009optomechanics,%
liu2011high}
and small levitated particles in optical cavities~\cite{%
chang2010cavity,%
gonzalez2019theory,%
millen2015cavity,%
martinetz2020quantum%
}.
In that configuration, 
the membrane vibration and the centre of mass motion and/or the wobbling motion of the particle provide mechanical degrees of freedom instead.

However, 
to the best of our knowledge,
there is neither theoretical nor experimental work on coupling between mechanical twist in a crystal and an optical field in optomechanical systems,
which would be a quantum optical counterpart of radiation torque discovered by Beth~\cite{beth1935direct,beth1936mechanical}.
With such an angular coupling between optical and mechanical subsystems,
the optomechanical system can acquire mechanical angular momentum and hence talk to other excitations with angular momenta (or magnetism) such as electron spin which provide emerging fields of spintronics. 
Indeed, mechanical angular momenta have been utilised to manipulate spins in various systems:
micromechanics~\cite{wallis2006einstein,zolfagharkhani2008nanomechanical,kobayashi2017spin,harii2019spin},
microfluidics~\cite{takahashi2016spin,kazerooni2020electron,kazerooni2021electrical},
atomic nuclei~\cite{chudo2014observation,wood2017magnetic,chudo2021barnett},
ultracold quantum gases~\cite{kawaguchi2006einstein,gawryluk2007resonant},
and quark-gluon plasma~\cite{adamczyk2017global}. 
Bringing the angular coupling will connect optomechanics to such other fields.

\begin{figure}[htbp]
  \centering
  \includegraphics[width=.7\linewidth]{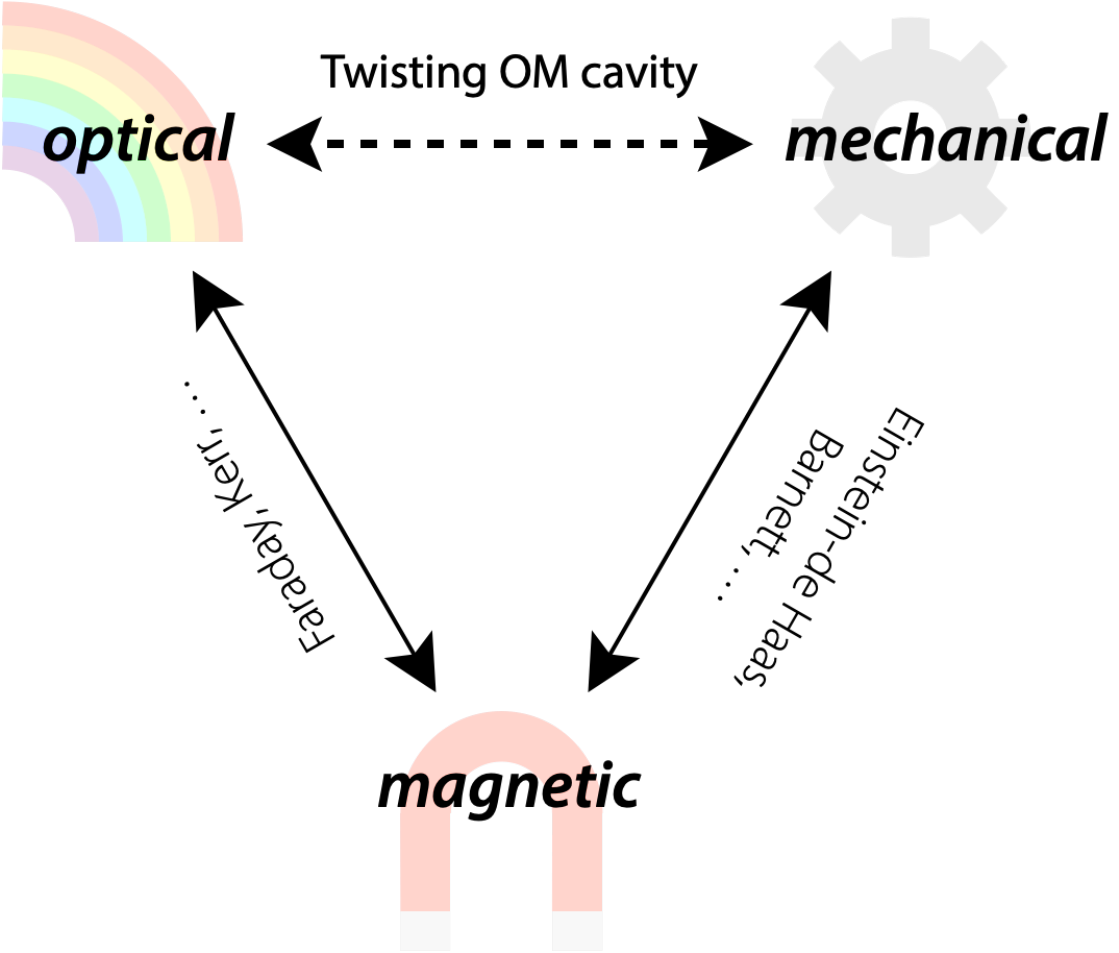}
  \caption{
    angular momentum transfer paths between optical, mechanical, and magnetic systems.
    Angular momenta can be exchanged between optical and magnetic systems, e.g., by Faraday and Kerr effects and between mechanical and magnetic systems, e.g., by Einstein-de Haas and Barnett effects.
    In this work, we elucidate an angular momentum transfer pathway between optical and mechanical systems (i.e.~twists in an optomechanical system).
  }
  \label{fig:am_transfer}
\end{figure}

In \figref{fig:am_transfer}, angular momentum transfer paths between optical, mechanical, and magnetic systems are illustrated.
Gyromagnetic effect (e.g., Einstein-de Haas and Barnett effects) is responsible for the angular momentum exchange between mechanical and magnetic systems and extensively utilised as reviewed above.
Magneto-optical effect (e.g., Faraday~\cite{faraday1846xli,faraday1846xlix,faraday1846xxvii} and Kerr~\cite{kerr1877xliii,kerr1878xxiv} effects) is another actor which enables angular momentum exchange beteen optical and magnetic degrees of freedom and has been utilised to analyse magnetic materials with high spatial and temporal resolutions~\cite{freeman2001advances,kirilyuk2010ultrafast,mccord2015progress,urs2016advanced,saidl2017optical,higo2018large} as well as to modulate light wave~\cite{shoji2014magneto,sun2016optical}.
We are going to make an angular momentum transfer pathway between optical and mechanical systems.

\begin{figure}[htbp]
  \centering
  \includegraphics[width=.9\linewidth]{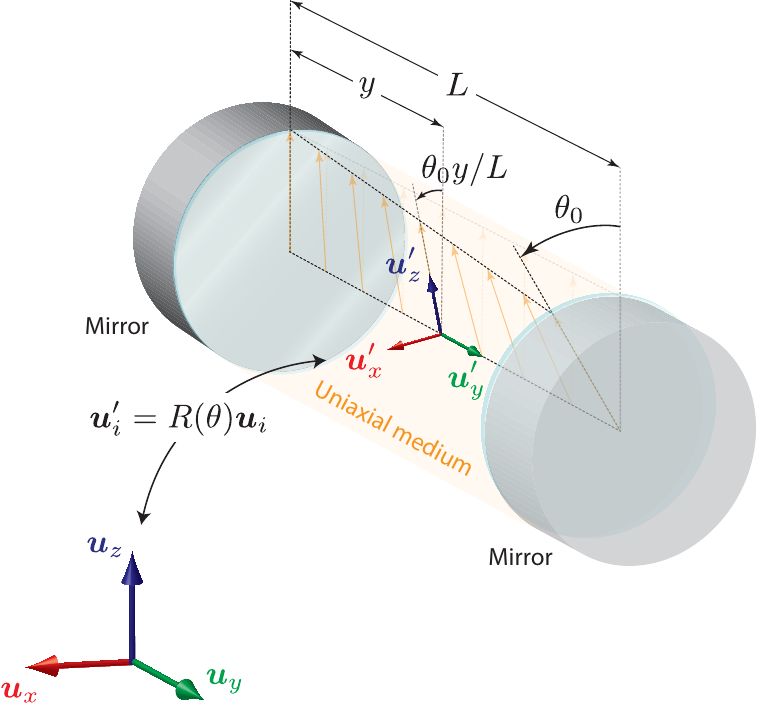}
  \caption{
    We consider a weakly twisted optical cavity in which an uniaxial medium is inserted.
    The angle of twist is $\theta_0$, and the cavity length is $L$.
    The optic axis of the uniaxial medium is indicated by orange arrows.
  }
  \label{fig:sketch}
\end{figure}

In this paper,
we derive an angular optomechanical interaction where two polarisation states in an optical cavity effectively behave as a two level system and interact with torsional mechanical motion of the cavity.
We consider an optical cavity filled with an anisotropic medium and twisting the cavity (see \figref{fig:sketch}).
The anisotropy breaks the rotational symmetry and plays a vital role to produce the angular optomechanical interaction.
Coordinate transformation associates the laboratory frame with a frame attached to the twisted cavity.
By utilising the transformation,
we can obtain the permittivity tensor in the laboratory frame and calculate the electromagnetic energy of the twisted cavity.
Quantisation of the total energy,
including the mechanical energy of the torsion,
yields our torsional optomechanical Hamiltonian.
We use the Gaussian unit and omit the time-dependence factor $e^{-i\omega t}$ for convenience throughout this paper.

\label{---permitivity tensor---}
\par \textit{Permittivity tensor of a twisted uniaxial medium.---}%
In the frame attached to the twisted cavity,
the permittivity tensor of the twisted uniaxial medium is 
\begin{align}
  \label{eq:epsilon_twisted}
  \epsilon' &= \epsilon_o \ux' \ux'^\intercal 
  + \epsilon_o \uy' \uy'^\intercal 
  + \epsilon_e \uz' \uz'^\intercal,
\end{align}
where $\bm{u}_i'$ is the unit vector in the $i'$ direction in the twisted frame.
In order to get the expression of the permittivity tensor in the laboratory frame,
we consider a coordinate transformation,
\begin{align}
  \label{eq:transformation}
  \bm{u}_i' &= R(\theta)\bm{u}_i\quad
  (i = x,y,z).
\end{align}
Here, $\bm{u}_i$ is the unit vector in the $i$ direction in the laboratory frame,
and we have used a rotation matrix,
\begin{align}
  \label{eq:rotation}
  R(\theta) &= 
  \begin{pmatrix}
    \cos \theta & 0 & \sin \theta\\
    0 & 1 & 0\\
    -\sin \theta & 0 & \cos \theta
  \end{pmatrix},
\end{align}
where $\theta = \theta_0 y / L$ is the twisted angle at the position $y$.
Substituting \eqref{eq:transformation} into \eqref{eq:epsilon_twisted},
we obtain the permittivity tensor in the laboratory frame,
\begin{align}
  \label{eq:epsilon_lab_nonapprox}
  \epsilon &= R(\theta) \epsilon' R^\intercal(\theta).
\end{align}
Up to the first order of $\theta$, we can expand
\begin{align}
  \label{eq:epsilon_lab}
  \epsilon &\simeq (\identity + i \theta \Lambda_5) \epsilon' (\identity - i \theta \Lambda_5)
           \simeq \epsilon' + \delta \epsilon \Lambda_4 \theta.
\end{align}
Here, $\identity \equiv \ux \ux^\intercal + \uy \uy^\intercal + \uz \uz^\intercal$ is the identity matrix,
$\Lambda_{4,5} = 
\begin{pmatrix}
  0 & 0 & 1\\
  0 & 0 & 0\\
  1 & 0 & 0
\end{pmatrix},\ 
\begin{pmatrix}
  0 & 0 & -i\\
  0 & 0 & 0\\
  i & 0 & 0
\end{pmatrix}$
are corresponding Gell-Mann matrices,
and we have set $\delta \epsilon \equiv \epsilon_e - \epsilon_o > 0$.
We regard $\delta \epsilon \Lambda_4 \theta$ as a perturbation.

\label{---eigenmode and eigenvalues ---}
\par \textit{Eigenmodes and eigenvalues.---}%
We derive eigenmodes and corresponding eigenvalues in the uniaxial cavity so as to calculate electromagnetic energy stored in the cavity.
From the Maxwell's curl equations~\cite{kong1986electromagnetic},
we can get
\begin{align}
  \label{eq:wave_eq}
  \left(\nabla \nabla^\intercal - \nabla^2 - \frac{\omega^2}{c^2} (\epsilon' + \delta \epsilon \Lambda_4 \theta) \right) \bm{E} &= 0,
\end{align}
where the unperturbed part is
\begin{align}
  \label{eq:wave_eq_unperturbed}
  \left(\nabla \nabla^\intercal - \nabla^2 - \frac{\omega^2}{c^2} \epsilon' \right) \bm{E} &= 0.
\end{align}
The perturbed part is responsible for the interaction between torsion and electromagnetic field.
We work in the reciprocal space and multiply the unperturbed equation by the inverse of the square root of the permittivity tensor from left-hand side to obtain an modified eigenvalue equation,
\begin{align}
  \label{eq:eigenvalue_equation}
  \epsilon'^{-\frac{1}{2}}\left( -\bm{k}\bm{k}^\intercal  + k^2 \identity \right)\epsilon'^{-\frac{1}{2}}  \bm{E}' &= \frac{\omega^2}{c^2} \bm{E}',\quad
 \bm{E}' = \epsilon'^{\frac{1}{2}} \bm{E}.
\end{align}
We can easily solve this Hermitian eigenvalue problem to obtain the eigenmodes and corresponding eigenvalues in the uniaxial cavity,
\begin{align}
  \label{eq:longitudinal}
  \bm{E}_l &= \frac{1}{\sqrt{V}}\epsilon'^{-\frac{1}{2}} \frac{\bm{k}'}{k'} e^{i\bm{k}\cdot\bm{r}},
  \quad 
  \left(\frac{\omega^2}{c^2}\right)_l = 0,
  \\
  \label{eq:ordinary}
  \bm{E}_o &= \frac{1}{\sqrt{V}}\epsilon'^{-\frac{1}{2}} \frac{\bm{k}' \times \uz}{k'} e^{i\bm{k}\cdot\bm{r}},
  \quad
  \left(\frac{\omega^2}{c^2}\right)_o = \frac{k^2}{\epsilon_o},
  \\
  \label{eq:extraordinary}
  \bm{E}_e &= \frac{1}{\sqrt{V}}\epsilon'^{-\frac{1}{2}} \frac{\bm{k}' \times \bm{k}' \times \uz}{k'k_\perp'} e^{i\bm{k}\cdot\bm{r}},
  \quad 
  \left(\frac{\omega^2}{c^2}\right)_e = \frac{k'^2}{\epsilon_o \epsilon_e},
\end{align}
where we have defined
\begin{align}
  \label{eq:wavevector}
  \bm{k}' \equiv \epsilon^{\frac{1}{2}} \bm{k}.
\end{align}
We have three eigenmodes: longitudinal, ordinary ray, and extraordinary ray modes.
The longitudinal mode always exists at zero frequency, and it does not contribute to the dynamics of the system.
The ordinary and extraordinary ray modes are well-known optical modes in uniaxial media~\cite{kong1986electromagnetic}.
The eigenmodes are orthogonal to each other because of the Hermiticity of the eigenvalue problem,
and we normalise the electric field by the volume of the system $V$ so that the mode functions satisfies,
$\int d\bm{r} \bm{E}_\sigma^\dagger \epsilon' \bm{E}_{\sigma'} = \delta_{\sigma \sigma'}$.

\label{---Torsional optomechanical Hamiltonian.---}
\par \textit{Quantisation.---}%
Here, we shall make a transition from classical to quantum theory.
By calculating the electromagnetic energy in the twisted cavity and by quantising it,
we can reach our torsional optomechanical Hamiltonian.
The electromagnetic energy is composed of electric and magnetic parts~\cite{jackson1999classical},
\begin{align}
  \label{eq:U}
  \mathcal{U} &= \frac{\mathrm{g}}{2} \int d\bm{r} \left( \bm{E}^\dagger \epsilon \bm{E} + \bm{H}^\dagger \mu \bm{H} \right),
\end{align}
where $\mathrm{g}=(8\pi)^{-1}$ is a Gaussian unit factor.
By substituting our permittivity tensor \eqref{eq:epsilon_lab} into \eqref{eq:U},
we get an `bare' electromagnetic energy,
which is the energy without torsion,
and an `interaction' energy between electromagnetic field and torsion,
\begin{align}
  \notag
  \mathcal{U} &= \frac{\mathrm{g}}{2} \int d\bm{r} \left( \bm{E}^\dagger \epsilon' \bm{E} + \bm{H}^\dagger \bm{H} \right) + \frac{\mathrm{g}}{2} \delta \epsilon \int \theta d\bm{r} \left( \bm{E}^\dagger \Lambda_4 \bm{E}\right),\\
  \label{eq:U_decomp}
              &\equiv \mathcal{U}_{\mathrm{em}} + \mathcal{U}_{\mathrm{int}}.
\end{align}
Here, we have set $\mu = \identity$.
We perform the eigenmode expansion of the electric field in the cavity and consider the two lowest energy modes,
\begin{align}
  \label{eq:E}
  \bm{E} = C_e \bm{E}_e + C_o \bm{E}_o,
\end{align}
where the wavevector associated with $\bm{E}_{o,e}$ is given by
$\bm{k}_{o,e} = ({\pi}/{L})\uy$,
while the frequency is 
\begin{align}
  \label{eq:omega_o,e}
  \omega_{o,e} &= \frac{ck}{n_o},\quad \frac{ck'}{n_o n_e},
\end{align}
respectively.
Here, we have defined $n_{o,e} = \sqrt{\epsilon_{o,e}}$.
We substitute \eqref{eq:E} into Eq.~\eqref{eq:U_decomp} to have the explicit expression for the interaction energy in our system,
\begin{align}
  \label{eq:U_int}
  \mathcal{U}_{\mathrm{int}} = -\frac{\mathrm{g}}{2}\left(\frac{1}{n_o^2} - \frac{1}{n_e^2}\right)\theta_0 (C_e^* C_o + C_o^* C_e).
\end{align}
Since we have already found the orthogonal eigenmodes,
we can quantise the electromagnetic field following the standard practice~\cite{cohen2004photon},
\begin{align}
  \label{eq:quantisation_em}
  \bm{E} = \sum_{\sigma=o,e} C_\sigma \bm{E}_\sigma &\mapsto \hat{\bm{E}} = i \sum_{\sigma=o,e} \sqrt{\frac{\hbar \omega_\sigma}{2}}\hat{a}_\sigma \bm{E}_\sigma,
\end{align}
where $\omega_\sigma$ is given in Eq.~\eqref{eq:omega_o,e},
and $\hat{a}_\sigma$ is the bosonic annihilation operator,
which annihilates a photon in the mode labeled by $\sigma$ and satisfies the commutation relation $[\hat{a}_\sigma, \hat{a}_{\sigma'}^\dagger]=\delta_{\sigma,\sigma'}$,

Since our medium filling the cavity produces a restoring force when twisted,
its dynamics may be modeled by a harmonic oscillator if the twist angle is small.
Thus, we can also quantise the torsional mechanical oscillation by following the standard quantisation scheme for harmonic oscillators~\cite{SchiffLeonardI1968Qm},
\begin{align}
  \label{eq:quantisation_m}
  \theta_0 &\mapsto \hat{\theta}_0 = \frac{1}{\sqrt{2}}(\hat{a}_m^\dagger + \hat{a}_m),
\end{align}
where we can write bosonic creation and annihilation operators, $
\hat{a}_m^\dagger = (\hat{Q} - i\hat{P})/\sqrt{2},\
\hat{a}_m = (\hat{Q} + i \hat{P})/\sqrt{2}
$
in terms of the position $\hat{Q}$ and canonical momentum $\hat{P}$.

\begin{figure}[tbp]
  \centering
  \includegraphics[width=.6\linewidth]{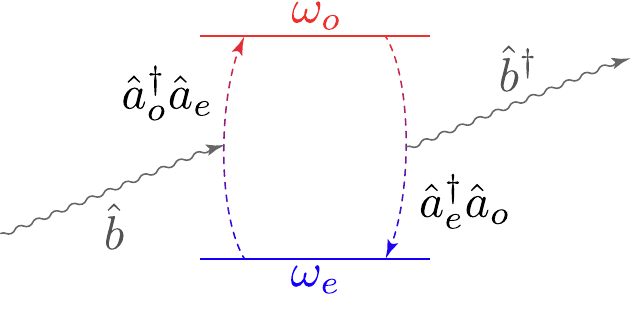}
  \caption{
    Energy level diagram of the torsional optomechanical Hamiltonian.
    The ordinary ray mode has higher energy than the extraordinary mode.
    Mediated by the twisting mechanical oscillation,
    they transit into each other and behave as an effective two level system coupled with a bosonic field.
  }
  \label{fig:energy_level}
\end{figure}

Substituting Eqs.~\eqref{eq:quantisation_em} and \eqref{eq:quantisation_m} into Eq.~\eqref{eq:U_int},
we can obtain the interaction Hamiltonian,
\begin{align}
  \label{eq:H_int}
  \hat{H}_{\mathrm{int}} 
  = \frac{-\hbar c}{16L\sqrt{2 n_o n_e}}
  \left(\frac{1}{n_o^2} - \frac{1}{n_e^2}\right)
  (\hat{a}_m^\dagger + \hat{a}_m) 
  (\hat{J}_+ + \hat{J}_-),
\end{align}
where we have defined $\hat{J}_+ = \hat{a}_o^\dagger \hat{a}_e$ and $\hat{J}_- = \hat{a}_e^\dagger \hat{a}_o$.
Those operators correspond to Schwinger's angular momentum operators,
which are composed of two kinds of bosons and satisfy commutation relations of the angular momentum type~\cite{schwinger1965on},
\begin{align}
  [\hat{J}_+, \hat{J}_-] = 2J_z,
  \quad
  [\hat{J}_z, \hat{J}_\pm] = \pm \hat{J}_\pm.
\end{align}
where we have defined 
$\hat{J}_z = {(\hat{a}_o^\dagger \hat{a}_o - \hat{a}_e^\dagger \hat{a}_e)}/{2}$.
This is why the interaction \eqref{eq:H_int} can be viewed as the interconversion of the kinetic angular momentum in the mechanical subsystem and the Schwinger angular momentum in the optical subsystems.
It can also be described by the energy level diagram shown in \figref{fig:energy_level}.
The extraordinary ray mode can transit into the ordinary ray mode by absorbing a torsional oscillation quantum,
and the ordinary mode relaxes to the extraordinary mode emitting a torsional oscillation quantum.
A similar interaction has been demonstrated in coupled resonator systems \cite{grudinin2010phonon,bahl2011stimulated,xu2014squeezing,sun2017phase,abudi2021resonators}, 
while their mechanical modes are not torsional unlike ours.
Although the two optical modes are linearly polarised and do not have any angular momentum,
they are still at the antipodal points on the Poincar\'{e} (Bloch) sphere and can be referred to as pseudospins as in the contexts of two-dimensional materials~\cite{schaibley2016valleytronics,kim2017graphene} and cold atomic gases~\cite{bloch2008many,zhai2015degenerate,enss2019universal}.
The interaction strength 
$g = \hbar c(1/n_o^2 - 1/n_e^2)/(16 L\sqrt{2 n_o n_e})$,
which appeared in the interaction Hamiltonian \eqref{eq:H_int},
is estimated to be of the order of
$\sim 0.1\ \mathrm{GHz}$
if we assume a quartz microcavity: 
$\epsilon_o = 2.25$;
$\epsilon_e = 1.001 \epsilon _ o$;
$L = 1\ \mathrm{\mu m}$.
Our torsional optomechanical Hamiltonian \eqref{eq:H_int} is different from the conventional one in that ours involves three kinds of bosons (ordinary and extraordinary rays and mechanical torsional oscillation) while the conventional one is composed of two bosons and proportional to the photon number.

We can also write the bare Hamiltonians of the electromagnetic field and mechanical oscillation,
\begin{align}
  \label{eq:bare_Hamiltonian}
  \hat{H}_{\mathrm{em}} = \sum_{\sigma=o,e} \hbar \omega_\sigma \hat{a}_\sigma^\dagger \hat{a}_\sigma,
  \quad
  \hat{H}_{\mathrm{mech}} &= \hbar \omega_0 \hat{a}_m^\dagger \hat{a}_m,
\end{align}
where $\omega_0$ is the eigenfrequency of the mechanical oscillation.
For simplicity, we have ignored the vacuum fluctuation energy here.
Finally, we have the torsional optomechanical Hamiltonian by summing up all the contributions,
$\hat{H} = \hat{H}_{\mathrm{em}} + \hat{H}_{\mathrm{mech}} + \hat{H}_{\mathrm{int}}$.

\label{---torsional optomechanical response---}
\par
\begin{figure}[tbp]
  \centering
  \includegraphics[width=\linewidth]{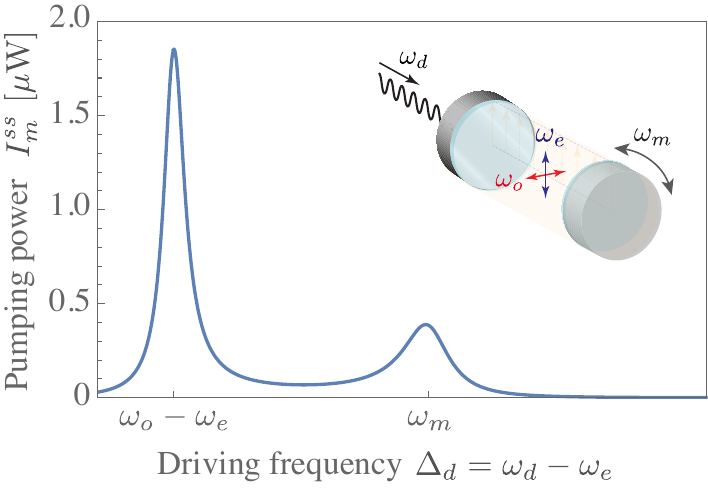}
  \caption{
    Torsional oscillation pumping by an external optical drive.
    The pumping spectrum is peaked at the mechanical frequency $\omega_m$ and the cavity photon (effective) frequency $\Delta = \omega_o - \omega_e$.
    To generate this figure,
    we use the following parameters:
    the optical mode frequency $ \Delta/2\pi = 5.0\ \mathrm{GHz}$
    and damping constant
    $\Gamma = 0.01\omega_m$;
    the mechanical mode frequency $\omega_m/2\pi = 5.5\ \mathrm{GHz}$
    and damping constant $\kappa = 0.01\omega_o$;
    the coupling constant \(g/2\pi = 62.4\ \mathrm{MHz}\);
    the temperature \(T = 300\ \mathrm{K}\);
    the input power \(P_\mathrm{in} = 100\ \mathrm{mW}\); 
    the permittivity ratio \(\epsilon_e/\epsilon_o = 1.001\).
    Inset: the schematic image of the configuration.
    The optomechanical cavity is modulated by an external drive with a frequency $\omega_d$,
    which produces torsional oscillation via the cavity photons.
  }
  \label{fig:pumping}
\end{figure}

 \textit{Torsional optomechanical response.---}%
Let us consider the response of the torsional optomecanical cavity under an external optical drive.
Here, we consider driving the ordinary ray mode.
Remind that we can selectively modulate the cavity modes by properly choosing the polarisation of the driving field because they are mutually orthogonal.
See the inset of \figref{fig:pumping} for the schematic of the configuration.
The driving effect can be taken into account by adding another Hamiltonian of the following form,
\begin{align}
H_\mathrm{drv} = 
h(t) \hat{a}_o(t) + h^*(t) \hat{a}_o^\dagger(t).
\end{align}
Note that the amplitude of the external drive is denoted by
\(h(t) = -i \hbar \Omega_d e^{i\omega_d t}\),
where 
\(\Omega_d = \sqrt{P_\mathrm{in}\kappa/(\hbar \omega_d)}\) 
is the coupling strength between the ordinary mode and the external drive with the driving power \(P_\mathrm{in}\), the cavity damping constant \(\kappa\), and the driving frequency $\omega_d$.
Note also that we can similarly study a mechanical drive by replacing the amplitude and operators with the machanical ones.
The detailed derivation of all equations appearing below is provided in the Supplemental Materials~\footnote{See Supplemental Material at \underline{URL inserted by publisher} for the derivation of equations.}.

Here, we shall study the energy changes in the mechanical subsystem under the drive,
which is one of the fundamental physical quantities of interest in our setup,
\begin{align}
  \label{eq:I_m}
  I_m (t) =
  -2\hbar \omega_m g 
  \operatorname{Im}
  \langle \hat{a}_m(t^+) \hat{J}_+(t^-) \rangle,
\end{align}
where $g$ is the interaction strength given in Eq.~\eqref{eq:H_int},
and the superscripts on the arguments,
$+$ and $-$,
denote the forward and backward branches on the Keldysh contour.
The statistical average is taken over the initial state,
\(\langle \hat{O} \rangle = \mathrm{tr}[\hat{\rho}_0 \hat{O} \hat{S}_C]\),
with the scattering operator,
$\hat{S}_C = \mathcal{T} \exp \int_C{\hat{H}_\mathrm{int}(t')}/{(i\hbar)}\mathrm{d}t'$,
stemming from the torsional optomechanical term, 
where $\mathcal{T}$ is the time ordering operator on the contour.

Expanding the scattering operator $\hat{S}_C$ in powers of the interaction strength $g$,
we perturbatively evaluate the statistical average \eqref{eq:I_m}.
Once we apply the Bloch-de Dominicis theorem,
each term becomes the product of the statistical averages of relevant operator pairs,
which are nothing but Green's functions.
At the nonequilibrium steady state under the external drive,
those products are reduced to convolutions.
In the frequency domain, we can write
\begin{align}
  I_m(t) \xrightarrow{t \rightarrow \infty} I_m^\mathrm{ss} 
  = 4 \hbar \omega_m \cdot (\hbar g)^2 \int 
  \mathrm{d}\omega
  \operatorname{Im} \chi_{\omega}^\mathfrak{R}
  \operatorname{Im} G_{\omega}^\mathfrak{R}
  \delta f_{\omega}
\end{align}
up to the second order in the torsional interaction.
Note that $\chi_\omega^\mathfrak{R}$ ($G_\omega^\mathfrak{R}$) is the retarded Green's function of the mechanical (optical) subsystem,
which gives the spectrum of each system,
and $\delta f_\omega$ corresponds to the population difference between the two subsystems at the nonequilibrium steady state.
This formula says that the torsional oscillation is pumped by the external drive via the nonequilibrium population difference,
and the pumping rate is proportional to the spectral overlap between the two subsystems.
Substituting the mechanical and optical spectra,
we can explicitly write
\begin{align}
  \frac{I_m^\mathrm{ss}}{4 \hbar \omega_m \cdot (\hbar g)^2} 
  = \operatorname{Im}
  \frac{1/\hbar}{\Delta_d - \omega_m + i\Gamma}
  \operatorname{Im}
  \frac{\langle a_e^\dagger a_e \rangle_0/\hbar}{\Delta_d - \Delta + i\kappa}
  \delta n_\mathrm{in},
\end{align}
where 
\(\Delta = \omega_o - \omega_e\ (\Delta_d = \omega_d - \omega_e)\)
is the ordinary ray (the external drive) frequency measured with reference to the extraordinary ray frequency,
and \(\delta n_\mathrm{in} = P_\mathrm{in}/(\hbar \omega_d)\) 
corresponds to the number of photons pumped into the cavity per unit time.
In \figref{fig:pumping}, the pumping rate spectrum is shown.
It is clear that the mechanical is peaked when the driving frequency coincides with the ordinary ray (effective) frequency $\Delta$ as well as the torsional oscillation frequency $\omega_m$.
The mechanical motion is resonantly pumped at those conditions.

\label{---conclusion---}
\par \textit{Conclusion.---}%
The frequency of light is far from that of mechanical oscillation,
and thus they do not interact with each other conventionally.
However, by introducing a birefringent crystal into a twistable cavity and hence breaking the rotational symmetry,
we can split the degenerated orthogonal linear polarisations,
which behave as an effective two level system,
and make it possible for them to interplay with the mechanical oscillation via permittivity tensor modulation by torsion.
We have quantised the electromagnetic energy density in the twisted birefringent cavity to identify the torsional optomechanical Hamiltonian.
The interaction term can be regarded as the interconversion of the kinetic angular momentum in the mechanical subsystem and the Schwinger angular momentum in the optical subsystem.
We have also revealed that torsional oscillation is resonantly pumped by modulating one of the two polarisations.
Our proposed scheme will twist optomechanical systems into devices which interconvert optical and mechanical angular momenta in arbitrary frequency range.

\begin{acknowledgments}
  D.~O.~and M.~M.~thank Yuya Ominato and Hiroyuki Tajima for fruitful discussions.
  This work was supported by the Priority Program of Chinese Academy of Sciences (Grant No.~XDB28000000) and JSPS KAKENHI (Grant Numbers: JP20H01863; JP21H04565; JP21H01800).
  D.~O.~is supported by the President's PhD Scholarships at Imperial College London and JSPS Overseas Research Fellowships.
\end{acknowledgments}

\bibliography{twist}

\begin{thebibliography}{68}%
\makeatletter
\providecommand \@ifxundefined [1]{%
 \@ifx{#1\undefined}
}%
\providecommand \@ifnum [1]{%
 \ifnum #1\expandafter \@firstoftwo
 \else \expandafter \@secondoftwo
 \fi
}%
\providecommand \@ifx [1]{%
 \ifx #1\expandafter \@firstoftwo
 \else \expandafter \@secondoftwo
 \fi
}%
\providecommand \natexlab [1]{#1}%
\providecommand \enquote  [1]{``#1''}%
\providecommand \bibnamefont  [1]{#1}%
\providecommand \bibfnamefont [1]{#1}%
\providecommand \citenamefont [1]{#1}%
\providecommand \href@noop [0]{\@secondoftwo}%
\providecommand \href [0]{\begingroup \@sanitize@url \@href}%
\providecommand \@href[1]{\@@startlink{#1}\@@href}%
\providecommand \@@href[1]{\endgroup#1\@@endlink}%
\providecommand \@sanitize@url [0]{\catcode `\\12\catcode `\$12\catcode
  `\&12\catcode `\#12\catcode `\^12\catcode `\_12\catcode `\%12\relax}%
\providecommand \@@startlink[1]{}%
\providecommand \@@endlink[0]{}%
\providecommand \url  [0]{\begingroup\@sanitize@url \@url }%
\providecommand \@url [1]{\endgroup\@href {#1}{\urlprefix }}%
\providecommand \urlprefix  [0]{URL }%
\providecommand \Eprint [0]{\href }%
\providecommand \doibase [0]{https://doi.org/}%
\providecommand \selectlanguage [0]{\@gobble}%
\providecommand \bibinfo  [0]{\@secondoftwo}%
\providecommand \bibfield  [0]{\@secondoftwo}%
\providecommand \translation [1]{[#1]}%
\providecommand \BibitemOpen [0]{}%
\providecommand \bibitemStop [0]{}%
\providecommand \bibitemNoStop [0]{.\EOS\space}%
\providecommand \EOS [0]{\spacefactor3000\relax}%
\providecommand \BibitemShut  [1]{\csname bibitem#1\endcsname}%
\let\auto@bib@innerbib\@empty
%</preamble>
\bibitem [{\citenamefont {Nichols}\ and\ \citenamefont
  {Hull}(1903{\natexlab{a}})}]{nichols1903pressure}%
  \BibitemOpen
  \bibfield  {author} {\bibinfo {author} {\bibfnamefont {E.~F.}\ \bibnamefont
  {Nichols}}\ and\ \bibinfo {author} {\bibfnamefont {G.~F.}\ \bibnamefont
  {Hull}},\ }\bibfield  {title} {\bibinfo {title} {The pressure due to
  radiation},\ }\href@noop {} {\bibfield  {journal} {\bibinfo  {journal}
  {Proceedings of the American Academy of Arts and Sciences}\ }\textbf
  {\bibinfo {volume} {38}},\ \bibinfo {pages} {559} (\bibinfo {year}
  {1903}{\natexlab{a}})}\BibitemShut {NoStop}%
\bibitem [{\citenamefont {Nichols}\ and\ \citenamefont
  {Hull}(1903{\natexlab{b}})}]{nichols1903pressure2nd}%
  \BibitemOpen
  \bibfield  {author} {\bibinfo {author} {\bibfnamefont {E.~F.}\ \bibnamefont
  {Nichols}}\ and\ \bibinfo {author} {\bibfnamefont {G.~F.}\ \bibnamefont
  {Hull}},\ }\bibfield  {title} {\bibinfo {title} {The pressure due to
  radiation.(second paper.)},\ }\href@noop {} {\bibfield  {journal} {\bibinfo
  {journal} {Physical Review (Series I)}\ }\textbf {\bibinfo {volume} {17}},\
  \bibinfo {pages} {26} (\bibinfo {year} {1903}{\natexlab{b}})}\BibitemShut
  {NoStop}%
\bibitem [{\citenamefont {Ashkin}(1970)}]{ashkin1970acceleration}%
  \BibitemOpen
  \bibfield  {author} {\bibinfo {author} {\bibfnamefont {A.}~\bibnamefont
  {Ashkin}},\ }\bibfield  {title} {\bibinfo {title} {Acceleration and trapping
  of particles by radiation pressure},\ }\href@noop {} {\bibfield  {journal}
  {\bibinfo  {journal} {Physical Review Letters}\ }\textbf {\bibinfo {volume}
  {24}},\ \bibinfo {pages} {156} (\bibinfo {year} {1970})}\BibitemShut
  {NoStop}%
\bibitem [{\citenamefont {Ashkin}\ \emph {et~al.}(1986)\citenamefont {Ashkin},
  \citenamefont {Dziedzic}, \citenamefont {Bjorkholm},\ and\ \citenamefont
  {Chu}}]{ashkin1986observation}%
  \BibitemOpen
  \bibfield  {author} {\bibinfo {author} {\bibfnamefont {A.}~\bibnamefont
  {Ashkin}}, \bibinfo {author} {\bibfnamefont {J.~M.}\ \bibnamefont
  {Dziedzic}}, \bibinfo {author} {\bibfnamefont {J.}~\bibnamefont
  {Bjorkholm}},\ and\ \bibinfo {author} {\bibfnamefont {S.}~\bibnamefont
  {Chu}},\ }\bibfield  {title} {\bibinfo {title} {Observation of a single-beam
  gradient force optical trap for dielectric particles},\ }\href@noop {}
  {\bibfield  {journal} {\bibinfo  {journal} {Optics letters}\ }\textbf
  {\bibinfo {volume} {11}},\ \bibinfo {pages} {288} (\bibinfo {year}
  {1986})}\BibitemShut {NoStop}%
\bibitem [{\citenamefont {Gao}\ \emph {et~al.}(2017)\citenamefont {Gao},
  \citenamefont {Ding}, \citenamefont {Nieto-Vesperinas}, \citenamefont {Ding},
  \citenamefont {Rahman}, \citenamefont {Zhang}, \citenamefont {Lim},\ and\
  \citenamefont {Qiu}}]{gao2017optical}%
  \BibitemOpen
  \bibfield  {author} {\bibinfo {author} {\bibfnamefont {D.}~\bibnamefont
  {Gao}}, \bibinfo {author} {\bibfnamefont {W.}~\bibnamefont {Ding}}, \bibinfo
  {author} {\bibfnamefont {M.}~\bibnamefont {Nieto-Vesperinas}}, \bibinfo
  {author} {\bibfnamefont {X.}~\bibnamefont {Ding}}, \bibinfo {author}
  {\bibfnamefont {M.}~\bibnamefont {Rahman}}, \bibinfo {author} {\bibfnamefont
  {T.}~\bibnamefont {Zhang}}, \bibinfo {author} {\bibfnamefont
  {C.}~\bibnamefont {Lim}},\ and\ \bibinfo {author} {\bibfnamefont {C.-W.}\
  \bibnamefont {Qiu}},\ }\bibfield  {title} {\bibinfo {title} {Optical
  manipulation from the microscale to the nanoscale: fundamentals, advances and
  prospects},\ }\href@noop {} {\bibfield  {journal} {\bibinfo  {journal}
  {Light: Science \& Applications}\ }\textbf {\bibinfo {volume} {6}},\ \bibinfo
  {pages} {e17039} (\bibinfo {year} {2017})}\BibitemShut {NoStop}%
\bibitem [{\citenamefont {Aspelmeyer}\ \emph {et~al.}(2014)\citenamefont
  {Aspelmeyer}, \citenamefont {Kippenberg},\ and\ \citenamefont
  {Marquardt}}]{aspelmeyer2014cavity}%
  \BibitemOpen
  \bibfield  {author} {\bibinfo {author} {\bibfnamefont {M.}~\bibnamefont
  {Aspelmeyer}}, \bibinfo {author} {\bibfnamefont {T.~J.}\ \bibnamefont
  {Kippenberg}},\ and\ \bibinfo {author} {\bibfnamefont {F.}~\bibnamefont
  {Marquardt}},\ }\bibfield  {title} {\bibinfo {title} {Cavity optomechanics},\
  }\href@noop {} {\bibfield  {journal} {\bibinfo  {journal} {Reviews of Modern
  Physics}\ }\textbf {\bibinfo {volume} {86}},\ \bibinfo {pages} {1391}
  (\bibinfo {year} {2014})}\BibitemShut {NoStop}%
\bibitem [{\citenamefont {Hofer}\ \emph {et~al.}(2010)\citenamefont {Hofer},
  \citenamefont {Schliesser},\ and\ \citenamefont
  {Kippenberg}}]{hofer2010cavity}%
  \BibitemOpen
  \bibfield  {author} {\bibinfo {author} {\bibfnamefont {J.}~\bibnamefont
  {Hofer}}, \bibinfo {author} {\bibfnamefont {A.}~\bibnamefont {Schliesser}},\
  and\ \bibinfo {author} {\bibfnamefont {T.~J.}\ \bibnamefont {Kippenberg}},\
  }\bibfield  {title} {\bibinfo {title} {Cavity optomechanics with
  ultrahigh-{Q} crystalline microresonators},\ }\href@noop {} {\bibfield
  {journal} {\bibinfo  {journal} {Physical Review A}\ }\textbf {\bibinfo
  {volume} {82}},\ \bibinfo {pages} {031804} (\bibinfo {year}
  {2010})}\BibitemShut {NoStop}%
\bibitem [{\citenamefont {Ding}\ \emph {et~al.}(2011)\citenamefont {Ding},
  \citenamefont {Baker}, \citenamefont {Senellart}, \citenamefont {Lemaitre},
  \citenamefont {Ducci}, \citenamefont {Leo},\ and\ \citenamefont
  {Favero}}]{ding2011wavelength}%
  \BibitemOpen
  \bibfield  {author} {\bibinfo {author} {\bibfnamefont {L.}~\bibnamefont
  {Ding}}, \bibinfo {author} {\bibfnamefont {C.}~\bibnamefont {Baker}},
  \bibinfo {author} {\bibfnamefont {P.}~\bibnamefont {Senellart}}, \bibinfo
  {author} {\bibfnamefont {A.}~\bibnamefont {Lemaitre}}, \bibinfo {author}
  {\bibfnamefont {S.}~\bibnamefont {Ducci}}, \bibinfo {author} {\bibfnamefont
  {G.}~\bibnamefont {Leo}},\ and\ \bibinfo {author} {\bibfnamefont
  {I.}~\bibnamefont {Favero}},\ }\bibfield  {title} {\bibinfo {title}
  {Wavelength-sized {GaAs} optomechanical resonators with gigahertz
  frequency},\ }\href@noop {} {\bibfield  {journal} {\bibinfo  {journal}
  {Applied Physics Letters}\ }\textbf {\bibinfo {volume} {98}},\ \bibinfo
  {pages} {113108} (\bibinfo {year} {2011})}\BibitemShut {NoStop}%
\bibitem [{\citenamefont {Forstner}\ \emph {et~al.}(2012)\citenamefont
  {Forstner}, \citenamefont {Prams}, \citenamefont {Knittel}, \citenamefont
  {Van~Ooijen}, \citenamefont {Swaim}, \citenamefont {Harris}, \citenamefont
  {Szorkovszky}, \citenamefont {Bowen},\ and\ \citenamefont
  {Rubinsztein-Dunlop}}]{forstner2012cavity}%
  \BibitemOpen
  \bibfield  {author} {\bibinfo {author} {\bibfnamefont {S.}~\bibnamefont
  {Forstner}}, \bibinfo {author} {\bibfnamefont {S.}~\bibnamefont {Prams}},
  \bibinfo {author} {\bibfnamefont {J.}~\bibnamefont {Knittel}}, \bibinfo
  {author} {\bibfnamefont {E.}~\bibnamefont {Van~Ooijen}}, \bibinfo {author}
  {\bibfnamefont {J.}~\bibnamefont {Swaim}}, \bibinfo {author} {\bibfnamefont
  {G.}~\bibnamefont {Harris}}, \bibinfo {author} {\bibfnamefont
  {A.}~\bibnamefont {Szorkovszky}}, \bibinfo {author} {\bibfnamefont
  {W.}~\bibnamefont {Bowen}},\ and\ \bibinfo {author} {\bibfnamefont
  {H.}~\bibnamefont {Rubinsztein-Dunlop}},\ }\bibfield  {title} {\bibinfo
  {title} {Cavity optomechanical magnetometer},\ }\href@noop {} {\bibfield
  {journal} {\bibinfo  {journal} {Physical review letters}\ }\textbf {\bibinfo
  {volume} {108}},\ \bibinfo {pages} {120801} (\bibinfo {year}
  {2012})}\BibitemShut {NoStop}%
\bibitem [{\citenamefont {Verhagen}\ \emph {et~al.}(2012)\citenamefont
  {Verhagen}, \citenamefont {Del{\'e}glise}, \citenamefont {Weis},
  \citenamefont {Schliesser},\ and\ \citenamefont
  {Kippenberg}}]{verhagen2012quantum}%
  \BibitemOpen
  \bibfield  {author} {\bibinfo {author} {\bibfnamefont {E.}~\bibnamefont
  {Verhagen}}, \bibinfo {author} {\bibfnamefont {S.}~\bibnamefont
  {Del{\'e}glise}}, \bibinfo {author} {\bibfnamefont {S.}~\bibnamefont {Weis}},
  \bibinfo {author} {\bibfnamefont {A.}~\bibnamefont {Schliesser}},\ and\
  \bibinfo {author} {\bibfnamefont {T.~J.}\ \bibnamefont {Kippenberg}},\
  }\bibfield  {title} {\bibinfo {title} {Quantum-coherent coupling of a
  mechanical oscillator to an optical cavity mode},\ }\href@noop {} {\bibfield
  {journal} {\bibinfo  {journal} {Nature}\ }\textbf {\bibinfo {volume} {482}},\
  \bibinfo {pages} {63} (\bibinfo {year} {2012})}\BibitemShut {NoStop}%
\bibitem [{\citenamefont {Chan}\ \emph {et~al.}(2011)\citenamefont {Chan},
  \citenamefont {Alegre}, \citenamefont {Safavi-Naeini}, \citenamefont {Hill},
  \citenamefont {Krause}, \citenamefont {Gr{\"o}blacher}, \citenamefont
  {Aspelmeyer},\ and\ \citenamefont {Painter}}]{chan2011laser}%
  \BibitemOpen
  \bibfield  {author} {\bibinfo {author} {\bibfnamefont {J.}~\bibnamefont
  {Chan}}, \bibinfo {author} {\bibfnamefont {T.}~\bibnamefont {Alegre}},
  \bibinfo {author} {\bibfnamefont {A.~H.}\ \bibnamefont {Safavi-Naeini}},
  \bibinfo {author} {\bibfnamefont {J.~T.}\ \bibnamefont {Hill}}, \bibinfo
  {author} {\bibfnamefont {A.}~\bibnamefont {Krause}}, \bibinfo {author}
  {\bibfnamefont {S.}~\bibnamefont {Gr{\"o}blacher}}, \bibinfo {author}
  {\bibfnamefont {M.}~\bibnamefont {Aspelmeyer}},\ and\ \bibinfo {author}
  {\bibfnamefont {O.}~\bibnamefont {Painter}},\ }\bibfield  {title} {\bibinfo
  {title} {Laser cooling of a nanomechanical oscillator into its quantum ground
  state},\ }\href@noop {} {\bibfield  {journal} {\bibinfo  {journal} {Nature}\
  }\textbf {\bibinfo {volume} {478}},\ \bibinfo {pages} {89} (\bibinfo {year}
  {2011})}\BibitemShut {NoStop}%
\bibitem [{\citenamefont {Eichenfield}\ \emph
  {et~al.}(2009{\natexlab{a}})\citenamefont {Eichenfield}, \citenamefont
  {Camacho}, \citenamefont {Chan}, \citenamefont {Vahala},\ and\ \citenamefont
  {Painter}}]{eichenfield2009picogram}%
  \BibitemOpen
  \bibfield  {author} {\bibinfo {author} {\bibfnamefont {M.}~\bibnamefont
  {Eichenfield}}, \bibinfo {author} {\bibfnamefont {R.}~\bibnamefont
  {Camacho}}, \bibinfo {author} {\bibfnamefont {J.}~\bibnamefont {Chan}},
  \bibinfo {author} {\bibfnamefont {K.~J.}\ \bibnamefont {Vahala}},\ and\
  \bibinfo {author} {\bibfnamefont {O.}~\bibnamefont {Painter}},\ }\bibfield
  {title} {\bibinfo {title} {A picogram-and nanometre-scale photonic-crystal
  optomechanical cavity},\ }\href@noop {} {\bibfield  {journal} {\bibinfo
  {journal} {nature}\ }\textbf {\bibinfo {volume} {459}},\ \bibinfo {pages}
  {550} (\bibinfo {year} {2009}{\natexlab{a}})}\BibitemShut {NoStop}%
\bibitem [{\citenamefont {Gavartin}\ \emph {et~al.}(2011)\citenamefont
  {Gavartin}, \citenamefont {Braive}, \citenamefont {Sagnes}, \citenamefont
  {Arcizet}, \citenamefont {Beveratos}, \citenamefont {Kippenberg},\ and\
  \citenamefont {Robert-Philip}}]{gavartin2011optomechanical}%
  \BibitemOpen
  \bibfield  {author} {\bibinfo {author} {\bibfnamefont {E.}~\bibnamefont
  {Gavartin}}, \bibinfo {author} {\bibfnamefont {R.}~\bibnamefont {Braive}},
  \bibinfo {author} {\bibfnamefont {I.}~\bibnamefont {Sagnes}}, \bibinfo
  {author} {\bibfnamefont {O.}~\bibnamefont {Arcizet}}, \bibinfo {author}
  {\bibfnamefont {A.}~\bibnamefont {Beveratos}}, \bibinfo {author}
  {\bibfnamefont {T.~J.}\ \bibnamefont {Kippenberg}},\ and\ \bibinfo {author}
  {\bibfnamefont {I.}~\bibnamefont {Robert-Philip}},\ }\bibfield  {title}
  {\bibinfo {title} {Optomechanical coupling in a two-dimensional photonic
  crystal defect cavity},\ }\href@noop {} {\bibfield  {journal} {\bibinfo
  {journal} {Physical review letters}\ }\textbf {\bibinfo {volume} {106}},\
  \bibinfo {pages} {203902} (\bibinfo {year} {2011})}\BibitemShut {NoStop}%
\bibitem [{\citenamefont {Maldovan}\ and\ \citenamefont
  {Thomas}(2006)}]{maldovan2006simultaneous}%
  \BibitemOpen
  \bibfield  {author} {\bibinfo {author} {\bibfnamefont {M.}~\bibnamefont
  {Maldovan}}\ and\ \bibinfo {author} {\bibfnamefont {E.~L.}\ \bibnamefont
  {Thomas}},\ }\bibfield  {title} {\bibinfo {title} {Simultaneous localization
  of photons and phonons in two-dimensional periodic structures},\ }\href@noop
  {} {\bibfield  {journal} {\bibinfo  {journal} {Applied Physics Letters}\
  }\textbf {\bibinfo {volume} {88}},\ \bibinfo {pages} {251907} (\bibinfo
  {year} {2006})}\BibitemShut {NoStop}%
\bibitem [{\citenamefont {Safavi-Naeini}\ and\ \citenamefont
  {Painter}(2010)}]{safavi2010design}%
  \BibitemOpen
  \bibfield  {author} {\bibinfo {author} {\bibfnamefont {A.~H.}\ \bibnamefont
  {Safavi-Naeini}}\ and\ \bibinfo {author} {\bibfnamefont {O.}~\bibnamefont
  {Painter}},\ }\bibfield  {title} {\bibinfo {title} {Design of optomechanical
  cavities and waveguides on a simultaneous bandgap phononic-photonic crystal
  slab},\ }\href@noop {} {\bibfield  {journal} {\bibinfo  {journal} {Optics
  express}\ }\textbf {\bibinfo {volume} {18}},\ \bibinfo {pages} {14926}
  (\bibinfo {year} {2010})}\BibitemShut {NoStop}%
\bibitem [{\citenamefont {Eichenfield}\ \emph
  {et~al.}(2009{\natexlab{b}})\citenamefont {Eichenfield}, \citenamefont
  {Chan}, \citenamefont {Camacho}, \citenamefont {Vahala},\ and\ \citenamefont
  {Painter}}]{eichenfield2009optomechanical}%
  \BibitemOpen
  \bibfield  {author} {\bibinfo {author} {\bibfnamefont {M.}~\bibnamefont
  {Eichenfield}}, \bibinfo {author} {\bibfnamefont {J.}~\bibnamefont {Chan}},
  \bibinfo {author} {\bibfnamefont {R.~M.}\ \bibnamefont {Camacho}}, \bibinfo
  {author} {\bibfnamefont {K.~J.}\ \bibnamefont {Vahala}},\ and\ \bibinfo
  {author} {\bibfnamefont {O.}~\bibnamefont {Painter}},\ }\bibfield  {title}
  {\bibinfo {title} {Optomechanical crystals},\ }\href@noop {} {\bibfield
  {journal} {\bibinfo  {journal} {Nature}\ }\textbf {\bibinfo {volume} {462}},\
  \bibinfo {pages} {78} (\bibinfo {year} {2009}{\natexlab{b}})}\BibitemShut
  {NoStop}%
\bibitem [{\citenamefont {Thompson}\ \emph {et~al.}(2008)\citenamefont
  {Thompson}, \citenamefont {Zwickl}, \citenamefont {Jayich}, \citenamefont
  {Marquardt}, \citenamefont {Girvin},\ and\ \citenamefont
  {Harris}}]{thompson2008strong}%
  \BibitemOpen
  \bibfield  {author} {\bibinfo {author} {\bibfnamefont {J.}~\bibnamefont
  {Thompson}}, \bibinfo {author} {\bibfnamefont {B.}~\bibnamefont {Zwickl}},
  \bibinfo {author} {\bibfnamefont {A.}~\bibnamefont {Jayich}}, \bibinfo
  {author} {\bibfnamefont {F.}~\bibnamefont {Marquardt}}, \bibinfo {author}
  {\bibfnamefont {S.}~\bibnamefont {Girvin}},\ and\ \bibinfo {author}
  {\bibfnamefont {J.}~\bibnamefont {Harris}},\ }\bibfield  {title} {\bibinfo
  {title} {Strong dispersive coupling of a high-finesse cavity to a
  micromechanical membrane},\ }\href@noop {} {\bibfield  {journal} {\bibinfo
  {journal} {Nature}\ }\textbf {\bibinfo {volume} {452}},\ \bibinfo {pages}
  {72} (\bibinfo {year} {2008})}\BibitemShut {NoStop}%
\bibitem [{\citenamefont {Wilson}\ \emph {et~al.}(2009)\citenamefont {Wilson},
  \citenamefont {Regal}, \citenamefont {Papp},\ and\ \citenamefont
  {Kimble}}]{wilson2009cavity}%
  \BibitemOpen
  \bibfield  {author} {\bibinfo {author} {\bibfnamefont {D.~J.}\ \bibnamefont
  {Wilson}}, \bibinfo {author} {\bibfnamefont {C.~A.}\ \bibnamefont {Regal}},
  \bibinfo {author} {\bibfnamefont {S.~B.}\ \bibnamefont {Papp}},\ and\
  \bibinfo {author} {\bibfnamefont {H.}~\bibnamefont {Kimble}},\ }\bibfield
  {title} {\bibinfo {title} {Cavity optomechanics with stoichiometric {SiN}
  films},\ }\href@noop {} {\bibfield  {journal} {\bibinfo  {journal} {Physical
  review letters}\ }\textbf {\bibinfo {volume} {103}},\ \bibinfo {pages}
  {207204} (\bibinfo {year} {2009})}\BibitemShut {NoStop}%
\bibitem [{\citenamefont {Favero}\ and\ \citenamefont
  {Karrai}(2009)}]{favero2009optomechanics}%
  \BibitemOpen
  \bibfield  {author} {\bibinfo {author} {\bibfnamefont {I.}~\bibnamefont
  {Favero}}\ and\ \bibinfo {author} {\bibfnamefont {K.}~\bibnamefont
  {Karrai}},\ }\bibfield  {title} {\bibinfo {title} {Optomechanics of
  deformable optical cavities},\ }\href@noop {} {\bibfield  {journal} {\bibinfo
   {journal} {Nature Photonics}\ }\textbf {\bibinfo {volume} {3}},\ \bibinfo
  {pages} {201} (\bibinfo {year} {2009})}\BibitemShut {NoStop}%
\bibitem [{\citenamefont {Liu}\ \emph {et~al.}(2011)\citenamefont {Liu},
  \citenamefont {Usami}, \citenamefont {Naesby}, \citenamefont {Bagci},
  \citenamefont {Polzik}, \citenamefont {Lodahl},\ and\ \citenamefont
  {Stobbe}}]{liu2011high}%
  \BibitemOpen
  \bibfield  {author} {\bibinfo {author} {\bibfnamefont {J.}~\bibnamefont
  {Liu}}, \bibinfo {author} {\bibfnamefont {K.}~\bibnamefont {Usami}}, \bibinfo
  {author} {\bibfnamefont {A.}~\bibnamefont {Naesby}}, \bibinfo {author}
  {\bibfnamefont {T.}~\bibnamefont {Bagci}}, \bibinfo {author} {\bibfnamefont
  {E.~S.}\ \bibnamefont {Polzik}}, \bibinfo {author} {\bibfnamefont
  {P.}~\bibnamefont {Lodahl}},\ and\ \bibinfo {author} {\bibfnamefont
  {S.}~\bibnamefont {Stobbe}},\ }\bibfield  {title} {\bibinfo {title} {High-{Q}
  optomechanical {GaAs} nanomembranes},\ }\href@noop {} {\bibfield  {journal}
  {\bibinfo  {journal} {Applied Physics Letters}\ }\textbf {\bibinfo {volume}
  {99}},\ \bibinfo {pages} {243102} (\bibinfo {year} {2011})}\BibitemShut
  {NoStop}%
\bibitem [{\citenamefont {Chang}\ \emph {et~al.}(2010)\citenamefont {Chang},
  \citenamefont {Regal}, \citenamefont {Papp}, \citenamefont {Wilson},
  \citenamefont {Ye}, \citenamefont {Painter}, \citenamefont {Kimble},\ and\
  \citenamefont {Zoller}}]{chang2010cavity}%
  \BibitemOpen
  \bibfield  {author} {\bibinfo {author} {\bibfnamefont {D.~E.}\ \bibnamefont
  {Chang}}, \bibinfo {author} {\bibfnamefont {C.}~\bibnamefont {Regal}},
  \bibinfo {author} {\bibfnamefont {S.}~\bibnamefont {Papp}}, \bibinfo {author}
  {\bibfnamefont {D.}~\bibnamefont {Wilson}}, \bibinfo {author} {\bibfnamefont
  {J.}~\bibnamefont {Ye}}, \bibinfo {author} {\bibfnamefont {O.}~\bibnamefont
  {Painter}}, \bibinfo {author} {\bibfnamefont {H.~J.}\ \bibnamefont
  {Kimble}},\ and\ \bibinfo {author} {\bibfnamefont {P.}~\bibnamefont
  {Zoller}},\ }\bibfield  {title} {\bibinfo {title} {Cavity opto-mechanics
  using an optically levitated nanosphere},\ }\href@noop {} {\bibfield
  {journal} {\bibinfo  {journal} {Proceedings of the National Academy of
  Sciences}\ }\textbf {\bibinfo {volume} {107}},\ \bibinfo {pages} {1005}
  (\bibinfo {year} {2010})}\BibitemShut {NoStop}%
\bibitem [{\citenamefont {Gonzalez-Ballestero}\ \emph
  {et~al.}(2019)\citenamefont {Gonzalez-Ballestero}, \citenamefont {Maurer},
  \citenamefont {Windey}, \citenamefont {Novotny}, \citenamefont {Reimann},\
  and\ \citenamefont {Romero-Isart}}]{gonzalez2019theory}%
  \BibitemOpen
  \bibfield  {author} {\bibinfo {author} {\bibfnamefont {C.}~\bibnamefont
  {Gonzalez-Ballestero}}, \bibinfo {author} {\bibfnamefont {P.}~\bibnamefont
  {Maurer}}, \bibinfo {author} {\bibfnamefont {D.}~\bibnamefont {Windey}},
  \bibinfo {author} {\bibfnamefont {L.}~\bibnamefont {Novotny}}, \bibinfo
  {author} {\bibfnamefont {R.}~\bibnamefont {Reimann}},\ and\ \bibinfo {author}
  {\bibfnamefont {O.}~\bibnamefont {Romero-Isart}},\ }\bibfield  {title}
  {\bibinfo {title} {Theory for cavity cooling of levitated nanoparticles via
  coherent scattering: {M}aster equation approach},\ }\href@noop {} {\bibfield
  {journal} {\bibinfo  {journal} {Physical Review A}\ }\textbf {\bibinfo
  {volume} {100}},\ \bibinfo {pages} {013805} (\bibinfo {year}
  {2019})}\BibitemShut {NoStop}%
\bibitem [{\citenamefont {Millen}\ \emph {et~al.}(2015)\citenamefont {Millen},
  \citenamefont {Fonseca}, \citenamefont {Mavrogordatos}, \citenamefont
  {Monteiro},\ and\ \citenamefont {Barker}}]{millen2015cavity}%
  \BibitemOpen
  \bibfield  {author} {\bibinfo {author} {\bibfnamefont {J.}~\bibnamefont
  {Millen}}, \bibinfo {author} {\bibfnamefont {P.}~\bibnamefont {Fonseca}},
  \bibinfo {author} {\bibfnamefont {T.}~\bibnamefont {Mavrogordatos}}, \bibinfo
  {author} {\bibfnamefont {T.}~\bibnamefont {Monteiro}},\ and\ \bibinfo
  {author} {\bibfnamefont {P.}~\bibnamefont {Barker}},\ }\bibfield  {title}
  {\bibinfo {title} {Cavity cooling a single charged levitated nanosphere},\
  }\href@noop {} {\bibfield  {journal} {\bibinfo  {journal} {Physical Review
  Letters}\ }\textbf {\bibinfo {volume} {114}},\ \bibinfo {pages} {123602}
  (\bibinfo {year} {2015})}\BibitemShut {NoStop}%
\bibitem [{\citenamefont {Martinetz}\ \emph {et~al.}(2020)\citenamefont
  {Martinetz}, \citenamefont {Hornberger}, \citenamefont {Millen},
  \citenamefont {Kim},\ and\ \citenamefont {Stickler}}]{martinetz2020quantum}%
  \BibitemOpen
  \bibfield  {author} {\bibinfo {author} {\bibfnamefont {L.}~\bibnamefont
  {Martinetz}}, \bibinfo {author} {\bibfnamefont {K.}~\bibnamefont
  {Hornberger}}, \bibinfo {author} {\bibfnamefont {J.}~\bibnamefont {Millen}},
  \bibinfo {author} {\bibfnamefont {M.}~\bibnamefont {Kim}},\ and\ \bibinfo
  {author} {\bibfnamefont {B.~A.}\ \bibnamefont {Stickler}},\ }\bibfield
  {title} {\bibinfo {title} {Quantum electromechanics with levitated
  nanoparticles},\ }\href@noop {} {\bibfield  {journal} {\bibinfo  {journal}
  {npj Quantum Information}\ }\textbf {\bibinfo {volume} {6}},\ \bibinfo
  {pages} {1} (\bibinfo {year} {2020})}\BibitemShut {NoStop}%
\bibitem [{\citenamefont {Beth}(1935)}]{beth1935direct}%
  \BibitemOpen
  \bibfield  {author} {\bibinfo {author} {\bibfnamefont {R.~A.}\ \bibnamefont
  {Beth}},\ }\bibfield  {title} {\bibinfo {title} {Direct detection of the
  angular momentum of light},\ }\href@noop {} {\bibfield  {journal} {\bibinfo
  {journal} {Physical Review}\ }\textbf {\bibinfo {volume} {48}},\ \bibinfo
  {pages} {471} (\bibinfo {year} {1935})}\BibitemShut {NoStop}%
\bibitem [{\citenamefont {Beth}(1936)}]{beth1936mechanical}%
  \BibitemOpen
  \bibfield  {author} {\bibinfo {author} {\bibfnamefont {R.~A.}\ \bibnamefont
  {Beth}},\ }\bibfield  {title} {\bibinfo {title} {Mechanical detection and
  measurement of the angular momentum of light},\ }\href@noop {} {\bibfield
  {journal} {\bibinfo  {journal} {Physical Review}\ }\textbf {\bibinfo {volume}
  {50}},\ \bibinfo {pages} {115} (\bibinfo {year} {1936})}\BibitemShut
  {NoStop}%
\bibitem [{\citenamefont {Wallis}\ \emph {et~al.}(2006)\citenamefont {Wallis},
  \citenamefont {Moreland},\ and\ \citenamefont {Kabos}}]{wallis2006einstein}%
  \BibitemOpen
  \bibfield  {author} {\bibinfo {author} {\bibfnamefont {T.~M.}\ \bibnamefont
  {Wallis}}, \bibinfo {author} {\bibfnamefont {J.}~\bibnamefont {Moreland}},\
  and\ \bibinfo {author} {\bibfnamefont {P.}~\bibnamefont {Kabos}},\ }\bibfield
   {title} {\bibinfo {title} {{Einstein}--{de Haas} effect in a {NiFe} film
  deposited on a microcantilever},\ }\href@noop {} {\bibfield  {journal}
  {\bibinfo  {journal} {Applied physics letters}\ }\textbf {\bibinfo {volume}
  {89}},\ \bibinfo {pages} {122502} (\bibinfo {year} {2006})}\BibitemShut
  {NoStop}%
\bibitem [{\citenamefont {Zolfagharkhani}\ \emph {et~al.}(2008)\citenamefont
  {Zolfagharkhani}, \citenamefont {Gaidarzhy}, \citenamefont {Degiovanni},
  \citenamefont {Kettemann}, \citenamefont {Fulde},\ and\ \citenamefont
  {Mohanty}}]{zolfagharkhani2008nanomechanical}%
  \BibitemOpen
  \bibfield  {author} {\bibinfo {author} {\bibfnamefont {G.}~\bibnamefont
  {Zolfagharkhani}}, \bibinfo {author} {\bibfnamefont {A.}~\bibnamefont
  {Gaidarzhy}}, \bibinfo {author} {\bibfnamefont {P.}~\bibnamefont
  {Degiovanni}}, \bibinfo {author} {\bibfnamefont {S.}~\bibnamefont
  {Kettemann}}, \bibinfo {author} {\bibfnamefont {P.}~\bibnamefont {Fulde}},\
  and\ \bibinfo {author} {\bibfnamefont {P.}~\bibnamefont {Mohanty}},\
  }\bibfield  {title} {\bibinfo {title} {Nanomechanical detection of itinerant
  electron spin flip},\ }\href@noop {} {\bibfield  {journal} {\bibinfo
  {journal} {Nature Nanotechnology}\ }\textbf {\bibinfo {volume} {3}},\
  \bibinfo {pages} {720} (\bibinfo {year} {2008})}\BibitemShut {NoStop}%
\bibitem [{\citenamefont {Kobayashi}\ \emph {et~al.}(2017)\citenamefont
  {Kobayashi}, \citenamefont {Yoshikawa}, \citenamefont {Matsuo}, \citenamefont
  {Iguchi}, \citenamefont {Maekawa}, \citenamefont {Saitoh},\ and\
  \citenamefont {Nozaki}}]{kobayashi2017spin}%
  \BibitemOpen
  \bibfield  {author} {\bibinfo {author} {\bibfnamefont {D.}~\bibnamefont
  {Kobayashi}}, \bibinfo {author} {\bibfnamefont {T.}~\bibnamefont
  {Yoshikawa}}, \bibinfo {author} {\bibfnamefont {M.}~\bibnamefont {Matsuo}},
  \bibinfo {author} {\bibfnamefont {R.}~\bibnamefont {Iguchi}}, \bibinfo
  {author} {\bibfnamefont {S.}~\bibnamefont {Maekawa}}, \bibinfo {author}
  {\bibfnamefont {E.}~\bibnamefont {Saitoh}},\ and\ \bibinfo {author}
  {\bibfnamefont {Y.}~\bibnamefont {Nozaki}},\ }\bibfield  {title} {\bibinfo
  {title} {Spin current generation using a surface acoustic wave generated via
  spin-rotation coupling},\ }\href@noop {} {\bibfield  {journal} {\bibinfo
  {journal} {Physical Review Letters}\ }\textbf {\bibinfo {volume} {119}},\
  \bibinfo {pages} {077202} (\bibinfo {year} {2017})}\BibitemShut {NoStop}%
\bibitem [{\citenamefont {Harii}\ \emph {et~al.}(2019)\citenamefont {Harii},
  \citenamefont {Seo}, \citenamefont {Tsutsumi}, \citenamefont {Chudo},
  \citenamefont {Oyanagi}, \citenamefont {Matsuo}, \citenamefont {Shiomi},
  \citenamefont {Ono}, \citenamefont {Maekawa},\ and\ \citenamefont
  {Saitoh}}]{harii2019spin}%
  \BibitemOpen
  \bibfield  {author} {\bibinfo {author} {\bibfnamefont {K.}~\bibnamefont
  {Harii}}, \bibinfo {author} {\bibfnamefont {Y.-J.}\ \bibnamefont {Seo}},
  \bibinfo {author} {\bibfnamefont {Y.}~\bibnamefont {Tsutsumi}}, \bibinfo
  {author} {\bibfnamefont {H.}~\bibnamefont {Chudo}}, \bibinfo {author}
  {\bibfnamefont {K.}~\bibnamefont {Oyanagi}}, \bibinfo {author} {\bibfnamefont
  {M.}~\bibnamefont {Matsuo}}, \bibinfo {author} {\bibfnamefont
  {Y.}~\bibnamefont {Shiomi}}, \bibinfo {author} {\bibfnamefont
  {T.}~\bibnamefont {Ono}}, \bibinfo {author} {\bibfnamefont {S.}~\bibnamefont
  {Maekawa}},\ and\ \bibinfo {author} {\bibfnamefont {E.}~\bibnamefont
  {Saitoh}},\ }\bibfield  {title} {\bibinfo {title} {Spin {Seebeck} mechanical
  force},\ }\href@noop {} {\bibfield  {journal} {\bibinfo  {journal} {Nature
  communications}\ }\textbf {\bibinfo {volume} {10}},\ \bibinfo {pages} {1}
  (\bibinfo {year} {2019})}\BibitemShut {NoStop}%
\bibitem [{\citenamefont {Takahashi}\ \emph {et~al.}(2016)\citenamefont
  {Takahashi}, \citenamefont {Matsuo}, \citenamefont {Ono}, \citenamefont
  {Harii}, \citenamefont {Chudo}, \citenamefont {Okayasu}, \citenamefont
  {Ieda}, \citenamefont {Takahashi}, \citenamefont {Maekawa},\ and\
  \citenamefont {Saitoh}}]{takahashi2016spin}%
  \BibitemOpen
  \bibfield  {author} {\bibinfo {author} {\bibfnamefont {R.}~\bibnamefont
  {Takahashi}}, \bibinfo {author} {\bibfnamefont {M.}~\bibnamefont {Matsuo}},
  \bibinfo {author} {\bibfnamefont {M.}~\bibnamefont {Ono}}, \bibinfo {author}
  {\bibfnamefont {K.}~\bibnamefont {Harii}}, \bibinfo {author} {\bibfnamefont
  {H.}~\bibnamefont {Chudo}}, \bibinfo {author} {\bibfnamefont
  {S.}~\bibnamefont {Okayasu}}, \bibinfo {author} {\bibfnamefont
  {J.}~\bibnamefont {Ieda}}, \bibinfo {author} {\bibfnamefont {S.}~\bibnamefont
  {Takahashi}}, \bibinfo {author} {\bibfnamefont {S.}~\bibnamefont {Maekawa}},\
  and\ \bibinfo {author} {\bibfnamefont {E.}~\bibnamefont {Saitoh}},\
  }\bibfield  {title} {\bibinfo {title} {Spin hydrodynamic generation},\
  }\href@noop {} {\bibfield  {journal} {\bibinfo  {journal} {Nature Physics}\
  }\textbf {\bibinfo {volume} {12}},\ \bibinfo {pages} {52} (\bibinfo {year}
  {2016})}\BibitemShut {NoStop}%
\bibitem [{\citenamefont {Kazerooni}\ \emph {et~al.}(2020)\citenamefont
  {Kazerooni}, \citenamefont {Thieme}, \citenamefont {Schumacher},\ and\
  \citenamefont {Cierpka}}]{kazerooni2020electron}%
  \BibitemOpen
  \bibfield  {author} {\bibinfo {author} {\bibfnamefont {H.~T.}\ \bibnamefont
  {Kazerooni}}, \bibinfo {author} {\bibfnamefont {A.}~\bibnamefont {Thieme}},
  \bibinfo {author} {\bibfnamefont {J.}~\bibnamefont {Schumacher}},\ and\
  \bibinfo {author} {\bibfnamefont {C.}~\bibnamefont {Cierpka}},\ }\bibfield
  {title} {\bibinfo {title} {Electron spin-vorticity coupling in pipe flows at
  low and high reynolds number},\ }\href@noop {} {\bibfield  {journal}
  {\bibinfo  {journal} {Physical Review Applied}\ }\textbf {\bibinfo {volume}
  {14}},\ \bibinfo {pages} {014002} (\bibinfo {year} {2020})}\BibitemShut
  {NoStop}%
\bibitem [{\citenamefont {Kazerooni}\ \emph {et~al.}(2021)\citenamefont
  {Kazerooni}, \citenamefont {Zinchenko}, \citenamefont {Schumacher},\ and\
  \citenamefont {Cierpka}}]{kazerooni2021electrical}%
  \BibitemOpen
  \bibfield  {author} {\bibinfo {author} {\bibfnamefont {H.~T.}\ \bibnamefont
  {Kazerooni}}, \bibinfo {author} {\bibfnamefont {G.}~\bibnamefont
  {Zinchenko}}, \bibinfo {author} {\bibfnamefont {J.}~\bibnamefont
  {Schumacher}},\ and\ \bibinfo {author} {\bibfnamefont {C.}~\bibnamefont
  {Cierpka}},\ }\bibfield  {title} {\bibinfo {title} {Electrical voltage by
  electron spin-vorticity coupling in laminar ducts},\ }\href@noop {}
  {\bibfield  {journal} {\bibinfo  {journal} {Physical Review Fluids}\ }\textbf
  {\bibinfo {volume} {6}},\ \bibinfo {pages} {043703} (\bibinfo {year}
  {2021})}\BibitemShut {NoStop}%
\bibitem [{\citenamefont {Chudo}\ \emph {et~al.}(2014)\citenamefont {Chudo},
  \citenamefont {Ono}, \citenamefont {Harii}, \citenamefont {Matsuo},
  \citenamefont {Ieda}, \citenamefont {Haruki}, \citenamefont {Okayasu},
  \citenamefont {Maekawa}, \citenamefont {Yasuoka},\ and\ \citenamefont
  {Saitoh}}]{chudo2014observation}%
  \BibitemOpen
  \bibfield  {author} {\bibinfo {author} {\bibfnamefont {H.}~\bibnamefont
  {Chudo}}, \bibinfo {author} {\bibfnamefont {M.}~\bibnamefont {Ono}}, \bibinfo
  {author} {\bibfnamefont {K.}~\bibnamefont {Harii}}, \bibinfo {author}
  {\bibfnamefont {M.}~\bibnamefont {Matsuo}}, \bibinfo {author} {\bibfnamefont
  {J.}~\bibnamefont {Ieda}}, \bibinfo {author} {\bibfnamefont {R.}~\bibnamefont
  {Haruki}}, \bibinfo {author} {\bibfnamefont {S.}~\bibnamefont {Okayasu}},
  \bibinfo {author} {\bibfnamefont {S.}~\bibnamefont {Maekawa}}, \bibinfo
  {author} {\bibfnamefont {H.}~\bibnamefont {Yasuoka}},\ and\ \bibinfo {author}
  {\bibfnamefont {E.}~\bibnamefont {Saitoh}},\ }\bibfield  {title} {\bibinfo
  {title} {Observation of {Barnett} fields in solids by nuclear magnetic
  resonance},\ }\href@noop {} {\bibfield  {journal} {\bibinfo  {journal}
  {Applied Physics Express}\ }\textbf {\bibinfo {volume} {7}},\ \bibinfo
  {pages} {063004} (\bibinfo {year} {2014})}\BibitemShut {NoStop}%
\bibitem [{\citenamefont {Wood}\ \emph {et~al.}(2017)\citenamefont {Wood},
  \citenamefont {Lilette}, \citenamefont {Fein}, \citenamefont {Perunicic},
  \citenamefont {Hollenberg}, \citenamefont {Scholten},\ and\ \citenamefont
  {Martin}}]{wood2017magnetic}%
  \BibitemOpen
  \bibfield  {author} {\bibinfo {author} {\bibfnamefont {A.}~\bibnamefont
  {Wood}}, \bibinfo {author} {\bibfnamefont {E.}~\bibnamefont {Lilette}},
  \bibinfo {author} {\bibfnamefont {Y.}~\bibnamefont {Fein}}, \bibinfo {author}
  {\bibfnamefont {V.}~\bibnamefont {Perunicic}}, \bibinfo {author}
  {\bibfnamefont {L.}~\bibnamefont {Hollenberg}}, \bibinfo {author}
  {\bibfnamefont {R.}~\bibnamefont {Scholten}},\ and\ \bibinfo {author}
  {\bibfnamefont {A.}~\bibnamefont {Martin}},\ }\bibfield  {title} {\bibinfo
  {title} {Magnetic pseudo-fields in a rotating electron--nuclear spin
  system},\ }\href@noop {} {\bibfield  {journal} {\bibinfo  {journal} {Nature
  Physics}\ }\textbf {\bibinfo {volume} {13}},\ \bibinfo {pages} {1070}
  (\bibinfo {year} {2017})}\BibitemShut {NoStop}%
\bibitem [{\citenamefont {Chudo}\ \emph {et~al.}(2021)\citenamefont {Chudo},
  \citenamefont {Matsuo}, \citenamefont {Maekawa},\ and\ \citenamefont
  {Saitoh}}]{chudo2021barnett}%
  \BibitemOpen
  \bibfield  {author} {\bibinfo {author} {\bibfnamefont {H.}~\bibnamefont
  {Chudo}}, \bibinfo {author} {\bibfnamefont {M.}~\bibnamefont {Matsuo}},
  \bibinfo {author} {\bibfnamefont {S.}~\bibnamefont {Maekawa}},\ and\ \bibinfo
  {author} {\bibfnamefont {E.}~\bibnamefont {Saitoh}},\ }\bibfield  {title}
  {\bibinfo {title} {Barnett field, rotational doppler effect, and berry phase
  studied by nuclear quadrupole resonance with rotation},\ }\href@noop {}
  {\bibfield  {journal} {\bibinfo  {journal} {Physical Review B}\ }\textbf
  {\bibinfo {volume} {103}},\ \bibinfo {pages} {174308} (\bibinfo {year}
  {2021})}\BibitemShut {NoStop}%
\bibitem [{\citenamefont {Kawaguchi}\ \emph {et~al.}(2006)\citenamefont
  {Kawaguchi}, \citenamefont {Saito},\ and\ \citenamefont
  {Ueda}}]{kawaguchi2006einstein}%
  \BibitemOpen
  \bibfield  {author} {\bibinfo {author} {\bibfnamefont {Y.}~\bibnamefont
  {Kawaguchi}}, \bibinfo {author} {\bibfnamefont {H.}~\bibnamefont {Saito}},\
  and\ \bibinfo {author} {\bibfnamefont {M.}~\bibnamefont {Ueda}},\ }\bibfield
  {title} {\bibinfo {title} {Einstein--de haas effect in dipolar bose-einstein
  condensates},\ }\href@noop {} {\bibfield  {journal} {\bibinfo  {journal}
  {Physical review letters}\ }\textbf {\bibinfo {volume} {96}},\ \bibinfo
  {pages} {080405} (\bibinfo {year} {2006})}\BibitemShut {NoStop}%
\bibitem [{\citenamefont {Gawryluk}\ \emph {et~al.}(2007)\citenamefont
  {Gawryluk}, \citenamefont {Brewczyk}, \citenamefont {Bongs},\ and\
  \citenamefont {Gajda}}]{gawryluk2007resonant}%
  \BibitemOpen
  \bibfield  {author} {\bibinfo {author} {\bibfnamefont {K.}~\bibnamefont
  {Gawryluk}}, \bibinfo {author} {\bibfnamefont {M.}~\bibnamefont {Brewczyk}},
  \bibinfo {author} {\bibfnamefont {K.}~\bibnamefont {Bongs}},\ and\ \bibinfo
  {author} {\bibfnamefont {M.}~\bibnamefont {Gajda}},\ }\bibfield  {title}
  {\bibinfo {title} {Resonant einstein--de haas effect in a rubidium
  condensate},\ }\href@noop {} {\bibfield  {journal} {\bibinfo  {journal}
  {Physical review letters}\ }\textbf {\bibinfo {volume} {99}},\ \bibinfo
  {pages} {130401} (\bibinfo {year} {2007})}\BibitemShut {NoStop}%
\bibitem [{\citenamefont {Adamczyk}\ \emph {et~al.}(2017)\citenamefont
  {Adamczyk}, \citenamefont {Adkins}, \citenamefont {Agakishiev}, \citenamefont
  {Aggarwal}, \citenamefont {Ahammed}, \citenamefont {Ajitanand}, \citenamefont
  {Alekseev}, \citenamefont {Anderson}, \citenamefont {Aoyama}, \citenamefont
  {Aparin} \emph {et~al.}}]{adamczyk2017global}%
  \BibitemOpen
  \bibfield  {author} {\bibinfo {author} {\bibfnamefont {L.}~\bibnamefont
  {Adamczyk}}, \bibinfo {author} {\bibfnamefont {J.}~\bibnamefont {Adkins}},
  \bibinfo {author} {\bibfnamefont {G.}~\bibnamefont {Agakishiev}}, \bibinfo
  {author} {\bibfnamefont {M.}~\bibnamefont {Aggarwal}}, \bibinfo {author}
  {\bibfnamefont {Z.}~\bibnamefont {Ahammed}}, \bibinfo {author} {\bibfnamefont
  {N.}~\bibnamefont {Ajitanand}}, \bibinfo {author} {\bibfnamefont
  {I.}~\bibnamefont {Alekseev}}, \bibinfo {author} {\bibfnamefont
  {D.}~\bibnamefont {Anderson}}, \bibinfo {author} {\bibfnamefont
  {R.}~\bibnamefont {Aoyama}}, \bibinfo {author} {\bibfnamefont
  {A.}~\bibnamefont {Aparin}}, \emph {et~al.},\ }\bibfield  {title} {\bibinfo
  {title} {Global $\lambda$ hyperon polarization in nuclear collisions},\
  }\href@noop {} {\bibfield  {journal} {\bibinfo  {journal} {Nature (London)}\
  }\textbf {\bibinfo {volume} {548}} (\bibinfo {year} {2017})}\BibitemShut
  {NoStop}%
\bibitem [{\citenamefont {Faraday}(1846{\natexlab{a}})}]{faraday1846xli}%
  \BibitemOpen
  \bibfield  {author} {\bibinfo {author} {\bibfnamefont {M.}~\bibnamefont
  {Faraday}},\ }\bibfield  {title} {\bibinfo {title} {{XLI}. {O}n the magnetic
  affection of light, and on the distinction between the ferromagnetic and
  diamagnetic conditions of matter},\ }\href@noop {} {\bibfield  {journal}
  {\bibinfo  {journal} {The London, Edinburgh, and Dublin Philosophical
  Magazine and Journal of Science}\ }\textbf {\bibinfo {volume} {29}},\
  \bibinfo {pages} {249} (\bibinfo {year} {1846}{\natexlab{a}})}\BibitemShut
  {NoStop}%
\bibitem [{\citenamefont {Faraday}(1846{\natexlab{b}})}]{faraday1846xlix}%
  \BibitemOpen
  \bibfield  {author} {\bibinfo {author} {\bibfnamefont {M.}~\bibnamefont
  {Faraday}},\ }\bibfield  {title} {\bibinfo {title} {{XLIX}. {Experimental}
  researches in electricity.---{Nineteenth} series},\ }\href@noop {} {\bibfield
   {journal} {\bibinfo  {journal} {The London, Edinburgh, and Dublin
  Philosophical Magazine and Journal of Science}\ }\textbf {\bibinfo {volume}
  {28}},\ \bibinfo {pages} {294} (\bibinfo {year}
  {1846}{\natexlab{b}})}\BibitemShut {NoStop}%
\bibitem [{\citenamefont {Faraday}(1846{\natexlab{c}})}]{faraday1846xxvii}%
  \BibitemOpen
  \bibfield  {author} {\bibinfo {author} {\bibfnamefont {M.}~\bibnamefont
  {Faraday}},\ }\bibfield  {title} {\bibinfo {title} {{XXVII}. {O}n the
  magnetic affection of light, and on the distinction between the ferromagnetic
  and diamagnetic conditions of matter},\ }\href@noop {} {\bibfield  {journal}
  {\bibinfo  {journal} {The London, Edinburgh, and Dublin Philosophical
  Magazine and Journal of Science}\ }\textbf {\bibinfo {volume} {29}},\
  \bibinfo {pages} {153} (\bibinfo {year} {1846}{\natexlab{c}})}\BibitemShut
  {NoStop}%
\bibitem [{\citenamefont {Kerr}(1877)}]{kerr1877xliii}%
  \BibitemOpen
  \bibfield  {author} {\bibinfo {author} {\bibfnamefont {J.}~\bibnamefont
  {Kerr}},\ }\bibfield  {title} {\bibinfo {title} {{XLIII}. {O}n rotation of
  the plane of polarization by reflection from the pole of a magnet},\
  }\href@noop {} {\bibfield  {journal} {\bibinfo  {journal} {The London,
  Edinburgh, and Dublin Philosophical Magazine and Journal of Science}\
  }\textbf {\bibinfo {volume} {3}},\ \bibinfo {pages} {321} (\bibinfo {year}
  {1877})}\BibitemShut {NoStop}%
\bibitem [{\citenamefont {Kerr}(1878)}]{kerr1878xxiv}%
  \BibitemOpen
  \bibfield  {author} {\bibinfo {author} {\bibfnamefont {J.}~\bibnamefont
  {Kerr}},\ }\bibfield  {title} {\bibinfo {title} {{XXIV}. {O}n reflection of
  polarized light from the equatorial surface of a magnet},\ }\href@noop {}
  {\bibfield  {journal} {\bibinfo  {journal} {The London, Edinburgh, and Dublin
  Philosophical Magazine and Journal of Science}\ }\textbf {\bibinfo {volume}
  {5}},\ \bibinfo {pages} {161} (\bibinfo {year} {1878})}\BibitemShut {NoStop}%
\bibitem [{\citenamefont {Freeman}\ and\ \citenamefont
  {Choi}(2001)}]{freeman2001advances}%
  \BibitemOpen
  \bibfield  {author} {\bibinfo {author} {\bibfnamefont {M.}~\bibnamefont
  {Freeman}}\ and\ \bibinfo {author} {\bibfnamefont {B.}~\bibnamefont {Choi}},\
  }\bibfield  {title} {\bibinfo {title} {Advances in magnetic microscopy},\
  }\href@noop {} {\bibfield  {journal} {\bibinfo  {journal} {Science}\ }\textbf
  {\bibinfo {volume} {294}},\ \bibinfo {pages} {1484} (\bibinfo {year}
  {2001})}\BibitemShut {NoStop}%
\bibitem [{\citenamefont {Kirilyuk}\ \emph {et~al.}(2010)\citenamefont
  {Kirilyuk}, \citenamefont {Kimel},\ and\ \citenamefont
  {Rasing}}]{kirilyuk2010ultrafast}%
  \BibitemOpen
  \bibfield  {author} {\bibinfo {author} {\bibfnamefont {A.}~\bibnamefont
  {Kirilyuk}}, \bibinfo {author} {\bibfnamefont {A.~V.}\ \bibnamefont
  {Kimel}},\ and\ \bibinfo {author} {\bibfnamefont {T.}~\bibnamefont
  {Rasing}},\ }\bibfield  {title} {\bibinfo {title} {Ultrafast optical
  manipulation of magnetic order},\ }\href@noop {} {\bibfield  {journal}
  {\bibinfo  {journal} {Reviews of Modern Physics}\ }\textbf {\bibinfo {volume}
  {82}},\ \bibinfo {pages} {2731} (\bibinfo {year} {2010})}\BibitemShut
  {NoStop}%
\bibitem [{\citenamefont {McCord}(2015)}]{mccord2015progress}%
  \BibitemOpen
  \bibfield  {author} {\bibinfo {author} {\bibfnamefont {J.}~\bibnamefont
  {McCord}},\ }\bibfield  {title} {\bibinfo {title} {Progress in magnetic
  domain observation by advanced magneto-optical microscopy},\ }\href@noop {}
  {\bibfield  {journal} {\bibinfo  {journal} {Journal of Physics D: Applied
  Physics}\ }\textbf {\bibinfo {volume} {48}},\ \bibinfo {pages} {333001}
  (\bibinfo {year} {2015})}\BibitemShut {NoStop}%
\bibitem [{\citenamefont {Urs}\ \emph {et~al.}(2016)\citenamefont {Urs},
  \citenamefont {Mozooni}, \citenamefont {Mazalski}, \citenamefont {Kustov},
  \citenamefont {Hayes}, \citenamefont {Deldar}, \citenamefont {Quandt},\ and\
  \citenamefont {McCord}}]{urs2016advanced}%
  \BibitemOpen
  \bibfield  {author} {\bibinfo {author} {\bibfnamefont {N.~O.}\ \bibnamefont
  {Urs}}, \bibinfo {author} {\bibfnamefont {B.}~\bibnamefont {Mozooni}},
  \bibinfo {author} {\bibfnamefont {P.}~\bibnamefont {Mazalski}}, \bibinfo
  {author} {\bibfnamefont {M.}~\bibnamefont {Kustov}}, \bibinfo {author}
  {\bibfnamefont {P.}~\bibnamefont {Hayes}}, \bibinfo {author} {\bibfnamefont
  {S.}~\bibnamefont {Deldar}}, \bibinfo {author} {\bibfnamefont
  {E.}~\bibnamefont {Quandt}},\ and\ \bibinfo {author} {\bibfnamefont
  {J.}~\bibnamefont {McCord}},\ }\bibfield  {title} {\bibinfo {title} {Advanced
  magneto-optical microscopy: Imaging from picoseconds to centimeters-imaging
  spin waves and temperature distributions},\ }\href@noop {} {\bibfield
  {journal} {\bibinfo  {journal} {AIP Advances}\ }\textbf {\bibinfo {volume}
  {6}},\ \bibinfo {pages} {055605} (\bibinfo {year} {2016})}\BibitemShut
  {NoStop}%
\bibitem [{\citenamefont {Saidl}\ \emph {et~al.}(2017)\citenamefont {Saidl},
  \citenamefont {N{\v{e}}mec}, \citenamefont {Wadley}, \citenamefont {Hills},
  \citenamefont {Campion}, \citenamefont {Nov{\'a}k}, \citenamefont {Edmonds},
  \citenamefont {Maccherozzi}, \citenamefont {Dhesi}, \citenamefont {Gallagher}
  \emph {et~al.}}]{saidl2017optical}%
  \BibitemOpen
  \bibfield  {author} {\bibinfo {author} {\bibfnamefont {V.}~\bibnamefont
  {Saidl}}, \bibinfo {author} {\bibfnamefont {P.}~\bibnamefont {N{\v{e}}mec}},
  \bibinfo {author} {\bibfnamefont {P.}~\bibnamefont {Wadley}}, \bibinfo
  {author} {\bibfnamefont {V.}~\bibnamefont {Hills}}, \bibinfo {author}
  {\bibfnamefont {R.}~\bibnamefont {Campion}}, \bibinfo {author} {\bibfnamefont
  {V.}~\bibnamefont {Nov{\'a}k}}, \bibinfo {author} {\bibfnamefont
  {K.}~\bibnamefont {Edmonds}}, \bibinfo {author} {\bibfnamefont
  {F.}~\bibnamefont {Maccherozzi}}, \bibinfo {author} {\bibfnamefont
  {S.}~\bibnamefont {Dhesi}}, \bibinfo {author} {\bibfnamefont
  {B.}~\bibnamefont {Gallagher}}, \emph {et~al.},\ }\bibfield  {title}
  {\bibinfo {title} {Optical determination of the n{\'e}el vector in a cumnas
  thin-film antiferromagnet},\ }\href@noop {} {\bibfield  {journal} {\bibinfo
  {journal} {Nature Photonics}\ }\textbf {\bibinfo {volume} {11}},\ \bibinfo
  {pages} {91} (\bibinfo {year} {2017})}\BibitemShut {NoStop}%
\bibitem [{\citenamefont {Higo}\ \emph {et~al.}(2018)\citenamefont {Higo},
  \citenamefont {Man}, \citenamefont {Gopman}, \citenamefont {Wu},
  \citenamefont {Koretsune}, \citenamefont {van't Erve}, \citenamefont
  {Kabanov}, \citenamefont {Rees}, \citenamefont {Li}, \citenamefont {Suzuki}
  \emph {et~al.}}]{higo2018large}%
  \BibitemOpen
  \bibfield  {author} {\bibinfo {author} {\bibfnamefont {T.}~\bibnamefont
  {Higo}}, \bibinfo {author} {\bibfnamefont {H.}~\bibnamefont {Man}}, \bibinfo
  {author} {\bibfnamefont {D.~B.}\ \bibnamefont {Gopman}}, \bibinfo {author}
  {\bibfnamefont {L.}~\bibnamefont {Wu}}, \bibinfo {author} {\bibfnamefont
  {T.}~\bibnamefont {Koretsune}}, \bibinfo {author} {\bibfnamefont {O.~M.}\
  \bibnamefont {van't Erve}}, \bibinfo {author} {\bibfnamefont {Y.~P.}\
  \bibnamefont {Kabanov}}, \bibinfo {author} {\bibfnamefont {D.}~\bibnamefont
  {Rees}}, \bibinfo {author} {\bibfnamefont {Y.}~\bibnamefont {Li}}, \bibinfo
  {author} {\bibfnamefont {M.-T.}\ \bibnamefont {Suzuki}}, \emph {et~al.},\
  }\bibfield  {title} {\bibinfo {title} {Large magneto-optical kerr effect and
  imaging of magnetic octupole domains in an antiferromagnetic metal},\
  }\href@noop {} {\bibfield  {journal} {\bibinfo  {journal} {Nature photonics}\
  }\textbf {\bibinfo {volume} {12}},\ \bibinfo {pages} {73} (\bibinfo {year}
  {2018})}\BibitemShut {NoStop}%
\bibitem [{\citenamefont {Shoji}\ and\ \citenamefont
  {Mizumoto}(2014)}]{shoji2014magneto}%
  \BibitemOpen
  \bibfield  {author} {\bibinfo {author} {\bibfnamefont {Y.}~\bibnamefont
  {Shoji}}\ and\ \bibinfo {author} {\bibfnamefont {T.}~\bibnamefont
  {Mizumoto}},\ }\bibfield  {title} {\bibinfo {title} {Magneto-optical
  non-reciprocal devices in silicon photonics},\ }\href@noop {} {\bibfield
  {journal} {\bibinfo  {journal} {Science and technology of advanced
  materials}\ }\textbf {\bibinfo {volume} {15}},\ \bibinfo {pages} {014602}
  (\bibinfo {year} {2014})}\BibitemShut {NoStop}%
\bibitem [{\citenamefont {Sun}\ \emph {et~al.}(2016)\citenamefont {Sun},
  \citenamefont {Martinez},\ and\ \citenamefont {Wang}}]{sun2016optical}%
  \BibitemOpen
  \bibfield  {author} {\bibinfo {author} {\bibfnamefont {Z.}~\bibnamefont
  {Sun}}, \bibinfo {author} {\bibfnamefont {A.}~\bibnamefont {Martinez}},\ and\
  \bibinfo {author} {\bibfnamefont {F.}~\bibnamefont {Wang}},\ }\bibfield
  {title} {\bibinfo {title} {Optical modulators with 2d layered materials},\
  }\href@noop {} {\bibfield  {journal} {\bibinfo  {journal} {Nature Photonics}\
  }\textbf {\bibinfo {volume} {10}},\ \bibinfo {pages} {227} (\bibinfo {year}
  {2016})}\BibitemShut {NoStop}%
\bibitem [{\citenamefont {Kong}(1986)}]{kong1986electromagnetic}%
  \BibitemOpen
  \bibfield  {author} {\bibinfo {author} {\bibfnamefont {J.~A.}\ \bibnamefont
  {Kong}},\ }\href@noop {} {\emph {\bibinfo {title} {Electromagnetic wave
  theory}}}\ (\bibinfo  {publisher} {John Wiley \& Sons, Inc., NewYork},\
  \bibinfo {year} {1986})\BibitemShut {NoStop}%
\bibitem [{\citenamefont {Jackson}(1999)}]{jackson1999classical}%
  \BibitemOpen
  \bibfield  {author} {\bibinfo {author} {\bibfnamefont {J.~D.}\ \bibnamefont
  {Jackson}},\ }\href@noop {} {\emph {\bibinfo {title} {Classical
  {E}lectrodynamics}}},\ \bibinfo {edition} {third edition}\ ed.\ (\bibinfo
  {publisher} {John Wiley \& Sons, Inc., NewYork},\ \bibinfo {year}
  {1999})\BibitemShut {NoStop}%
\bibitem [{\citenamefont {Cohen-Tannoudji}\ \emph {et~al.}(2004)\citenamefont
  {Cohen-Tannoudji}, \citenamefont {Dupont-Roc},\ and\ \citenamefont
  {Grynberg}}]{cohen2004photon}%
  \BibitemOpen
  \bibfield  {author} {\bibinfo {author} {\bibfnamefont {C.}~\bibnamefont
  {Cohen-Tannoudji}}, \bibinfo {author} {\bibfnamefont {J.}~\bibnamefont
  {Dupont-Roc}},\ and\ \bibinfo {author} {\bibfnamefont {G.}~\bibnamefont
  {Grynberg}},\ }\href@noop {} {\emph {\bibinfo {title} {{P}hotons and
  {A}toms}}}\ (\bibinfo  {publisher} {Wiley-VCH Verlag GmbH \& Co. KGaA,
  Weinheim},\ \bibinfo {year} {2004})\BibitemShut {NoStop}%
\bibitem [{\citenamefont {Schiff}(1968)}]{SchiffLeonardI1968Qm}%
  \BibitemOpen
  \bibfield  {author} {\bibinfo {author} {\bibfnamefont {L.~I.}\ \bibnamefont
  {Schiff}},\ }\href@noop {} {\emph {\bibinfo {title} {Quantum mechanics}}},\
  \bibinfo {edition} {3rd}\ ed.,\ International series in pure and applied
  physics\ (\bibinfo  {publisher} {McGraw-Hill},\ \bibinfo {address} {New York
  ; Maidenhead},\ \bibinfo {year} {1968})\BibitemShut {NoStop}%
\bibitem [{\citenamefont {Schwinger}\ \emph {et~al.}(1965)\citenamefont
  {Schwinger}, \citenamefont {Biedenharn},\ and\ \citenamefont
  {Van~Dam}}]{schwinger1965on}%
  \BibitemOpen
  \bibfield  {author} {\bibinfo {author} {\bibfnamefont {J.}~\bibnamefont
  {Schwinger}}, \bibinfo {author} {\bibfnamefont {L.}~\bibnamefont
  {Biedenharn}},\ and\ \bibinfo {author} {\bibfnamefont {H.}~\bibnamefont
  {Van~Dam}},\ }\href@noop {} {\emph {\bibinfo {title} {Quantum theory of
  angular momentum}}}\ (\bibinfo  {publisher} {Academic Press, New York},\
  \bibinfo {year} {1965})\BibitemShut {NoStop}%
\bibitem [{\citenamefont {Grudinin}\ \emph {et~al.}(2010)\citenamefont
  {Grudinin}, \citenamefont {Lee}, \citenamefont {Painter},\ and\ \citenamefont
  {Vahala}}]{grudinin2010phonon}%
  \BibitemOpen
  \bibfield  {author} {\bibinfo {author} {\bibfnamefont {I.~S.}\ \bibnamefont
  {Grudinin}}, \bibinfo {author} {\bibfnamefont {H.}~\bibnamefont {Lee}},
  \bibinfo {author} {\bibfnamefont {O.}~\bibnamefont {Painter}},\ and\ \bibinfo
  {author} {\bibfnamefont {K.~J.}\ \bibnamefont {Vahala}},\ }\bibfield  {title}
  {\bibinfo {title} {Phonon laser action in a tunable two-level system},\
  }\href@noop {} {\bibfield  {journal} {\bibinfo  {journal} {Physical review
  letters}\ }\textbf {\bibinfo {volume} {104}},\ \bibinfo {pages} {083901}
  (\bibinfo {year} {2010})}\BibitemShut {NoStop}%
\bibitem [{\citenamefont {Bahl}\ \emph {et~al.}(2011)\citenamefont {Bahl},
  \citenamefont {Zehnpfennig}, \citenamefont {Tomes},\ and\ \citenamefont
  {Carmon}}]{bahl2011stimulated}%
  \BibitemOpen
  \bibfield  {author} {\bibinfo {author} {\bibfnamefont {G.}~\bibnamefont
  {Bahl}}, \bibinfo {author} {\bibfnamefont {J.}~\bibnamefont {Zehnpfennig}},
  \bibinfo {author} {\bibfnamefont {M.}~\bibnamefont {Tomes}},\ and\ \bibinfo
  {author} {\bibfnamefont {T.}~\bibnamefont {Carmon}},\ }\bibfield  {title}
  {\bibinfo {title} {Stimulated optomechanical excitation of surface acoustic
  waves in a microdevice},\ }\href@noop {} {\bibfield  {journal} {\bibinfo
  {journal} {Nature communications}\ }\textbf {\bibinfo {volume} {2}},\
  \bibinfo {pages} {1} (\bibinfo {year} {2011})}\BibitemShut {NoStop}%
\bibitem [{\citenamefont {Xu}\ and\ \citenamefont
  {Taylor}(2014)}]{xu2014squeezing}%
  \BibitemOpen
  \bibfield  {author} {\bibinfo {author} {\bibfnamefont {X.}~\bibnamefont
  {Xu}}\ and\ \bibinfo {author} {\bibfnamefont {J.~M.}\ \bibnamefont
  {Taylor}},\ }\bibfield  {title} {\bibinfo {title} {Squeezing in a coupled
  two-mode optomechanical system for force sensing below the standard quantum
  limit},\ }\href@noop {} {\bibfield  {journal} {\bibinfo  {journal} {Physical
  Review A}\ }\textbf {\bibinfo {volume} {90}},\ \bibinfo {pages} {043848}
  (\bibinfo {year} {2014})}\BibitemShut {NoStop}%
\bibitem [{\citenamefont {Sun}\ \emph {et~al.}(2017)\citenamefont {Sun},
  \citenamefont {Mao}, \citenamefont {Dai}, \citenamefont {Ficek},
  \citenamefont {He},\ and\ \citenamefont {Gong}}]{sun2017phase}%
  \BibitemOpen
  \bibfield  {author} {\bibinfo {author} {\bibfnamefont {F.}~\bibnamefont
  {Sun}}, \bibinfo {author} {\bibfnamefont {D.}~\bibnamefont {Mao}}, \bibinfo
  {author} {\bibfnamefont {Y.}~\bibnamefont {Dai}}, \bibinfo {author}
  {\bibfnamefont {Z.}~\bibnamefont {Ficek}}, \bibinfo {author} {\bibfnamefont
  {Q.}~\bibnamefont {He}},\ and\ \bibinfo {author} {\bibfnamefont
  {Q.}~\bibnamefont {Gong}},\ }\bibfield  {title} {\bibinfo {title} {Phase
  control of entanglement and quantum steering in a three-mode optomechanical
  system},\ }\href@noop {} {\bibfield  {journal} {\bibinfo  {journal} {New
  Journal of Physics}\ }\textbf {\bibinfo {volume} {19}},\ \bibinfo {pages}
  {123039} (\bibinfo {year} {2017})}\BibitemShut {NoStop}%
\bibitem [{\citenamefont {Abudi}\ \emph {et~al.}(2021)\citenamefont {Abudi},
  \citenamefont {Douvidzon}, \citenamefont {Bathish},\ and\ \citenamefont
  {Carmon}}]{abudi2021resonators}%
  \BibitemOpen
  \bibfield  {author} {\bibinfo {author} {\bibfnamefont {T.~L.}\ \bibnamefont
  {Abudi}}, \bibinfo {author} {\bibfnamefont {M.}~\bibnamefont {Douvidzon}},
  \bibinfo {author} {\bibfnamefont {B.}~\bibnamefont {Bathish}},\ and\ \bibinfo
  {author} {\bibfnamefont {T.}~\bibnamefont {Carmon}},\ }\bibfield  {title}
  {\bibinfo {title} {Resonators made of a disk and a movable
  continuous-membrane},\ }\href@noop {} {\bibfield  {journal} {\bibinfo
  {journal} {APL Photonics}\ }\textbf {\bibinfo {volume} {6}},\ \bibinfo
  {pages} {036105} (\bibinfo {year} {2021})}\BibitemShut {NoStop}%
\bibitem [{\citenamefont {Schaibley}\ \emph {et~al.}(2016)\citenamefont
  {Schaibley}, \citenamefont {Yu}, \citenamefont {Clark}, \citenamefont
  {Rivera}, \citenamefont {Ross}, \citenamefont {Seyler}, \citenamefont {Yao},\
  and\ \citenamefont {Xu}}]{schaibley2016valleytronics}%
  \BibitemOpen
  \bibfield  {author} {\bibinfo {author} {\bibfnamefont {J.~R.}\ \bibnamefont
  {Schaibley}}, \bibinfo {author} {\bibfnamefont {H.}~\bibnamefont {Yu}},
  \bibinfo {author} {\bibfnamefont {G.}~\bibnamefont {Clark}}, \bibinfo
  {author} {\bibfnamefont {P.}~\bibnamefont {Rivera}}, \bibinfo {author}
  {\bibfnamefont {J.~S.}\ \bibnamefont {Ross}}, \bibinfo {author}
  {\bibfnamefont {K.~L.}\ \bibnamefont {Seyler}}, \bibinfo {author}
  {\bibfnamefont {W.}~\bibnamefont {Yao}},\ and\ \bibinfo {author}
  {\bibfnamefont {X.}~\bibnamefont {Xu}},\ }\bibfield  {title} {\bibinfo
  {title} {Valleytronics in 2d materials},\ }\href@noop {} {\bibfield
  {journal} {\bibinfo  {journal} {Nature Reviews Materials}\ }\textbf {\bibinfo
  {volume} {1}},\ \bibinfo {pages} {1} (\bibinfo {year} {2016})}\BibitemShut
  {NoStop}%
\bibitem [{\citenamefont {Kim}(2017)}]{kim2017graphene}%
  \BibitemOpen
  \bibfield  {author} {\bibinfo {author} {\bibfnamefont {P.}~\bibnamefont
  {Kim}},\ }\bibfield  {title} {\bibinfo {title} {Graphene and relativistic
  quantum physics},\ }in\ \href@noop {} {\emph {\bibinfo {booktitle} {Dirac
  Matter}}}\ (\bibinfo  {publisher} {Springer},\ \bibinfo {year} {2017})\ pp.\
  \bibinfo {pages} {1--23}\BibitemShut {NoStop}%
\bibitem [{\citenamefont {Bloch}\ \emph {et~al.}(2008)\citenamefont {Bloch},
  \citenamefont {Dalibard},\ and\ \citenamefont {Zwerger}}]{bloch2008many}%
  \BibitemOpen
  \bibfield  {author} {\bibinfo {author} {\bibfnamefont {I.}~\bibnamefont
  {Bloch}}, \bibinfo {author} {\bibfnamefont {J.}~\bibnamefont {Dalibard}},\
  and\ \bibinfo {author} {\bibfnamefont {W.}~\bibnamefont {Zwerger}},\
  }\bibfield  {title} {\bibinfo {title} {Many-body physics with ultracold
  gases},\ }\href@noop {} {\bibfield  {journal} {\bibinfo  {journal} {Reviews
  of modern physics}\ }\textbf {\bibinfo {volume} {80}},\ \bibinfo {pages}
  {885} (\bibinfo {year} {2008})}\BibitemShut {NoStop}%
\bibitem [{\citenamefont {Zhai}(2015)}]{zhai2015degenerate}%
  \BibitemOpen
  \bibfield  {author} {\bibinfo {author} {\bibfnamefont {H.}~\bibnamefont
  {Zhai}},\ }\bibfield  {title} {\bibinfo {title} {Degenerate quantum gases
  with spin--orbit coupling: a review},\ }\href@noop {} {\bibfield  {journal}
  {\bibinfo  {journal} {Reports on Progress in Physics}\ }\textbf {\bibinfo
  {volume} {78}},\ \bibinfo {pages} {026001} (\bibinfo {year}
  {2015})}\BibitemShut {NoStop}%
\bibitem [{\citenamefont {Enss}\ and\ \citenamefont
  {Thywissen}(2019)}]{enss2019universal}%
  \BibitemOpen
  \bibfield  {author} {\bibinfo {author} {\bibfnamefont {T.}~\bibnamefont
  {Enss}}\ and\ \bibinfo {author} {\bibfnamefont {J.~H.}\ \bibnamefont
  {Thywissen}},\ }\bibfield  {title} {\bibinfo {title} {Universal spin
  transport and quantum bounds for unitary fermions},\ }\href@noop {}
  {\bibfield  {journal} {\bibinfo  {journal} {Annual Review of Condensed Matter
  Physics}\ }\textbf {\bibinfo {volume} {10}},\ \bibinfo {pages} {85} (\bibinfo
  {year} {2019})}\BibitemShut {NoStop}%
\bibitem [{Note1()}]{Note1}%
  \BibitemOpen
  \bibinfo {note} {See Supplemental Material at \protect \underline {URL
  inserted by publisher} for the derivation of equations.}\BibitemShut {Stop}%
\end{thebibliography}%


\begin{thebibliography}{1}%
\makeatletter
\providecommand \@ifxundefined [1]{%
 \@ifx{#1\undefined}
}%
\providecommand \@ifnum [1]{%
 \ifnum #1\expandafter \@firstoftwo
 \else \expandafter \@secondoftwo
 \fi
}%
\providecommand \@ifx [1]{%
 \ifx #1\expandafter \@firstoftwo
 \else \expandafter \@secondoftwo
 \fi
}%
\providecommand \natexlab [1]{#1}%
\providecommand \enquote  [1]{``#1''}%
\providecommand \bibnamefont  [1]{#1}%
\providecommand \bibfnamefont [1]{#1}%
\providecommand \citenamefont [1]{#1}%
\providecommand \href@noop [0]{\@secondoftwo}%
\providecommand \href [0]{\begingroup \@sanitize@url \@href}%
\providecommand \@href[1]{\@@startlink{#1}\@@href}%
\providecommand \@@href[1]{\endgroup#1\@@endlink}%
\providecommand \@sanitize@url [0]{\catcode `\\12\catcode `\$12\catcode
  `\&12\catcode `\#12\catcode `\^12\catcode `\_12\catcode `\%12\relax}%
\providecommand \@@startlink[1]{}%
\providecommand \@@endlink[0]{}%
\providecommand \url  [0]{\begingroup\@sanitize@url \@url }%
\providecommand \@url [1]{\endgroup\@href {#1}{\urlprefix }}%
\providecommand \urlprefix  [0]{URL }%
\providecommand \Eprint [0]{\href }%
\providecommand \doibase [0]{https://doi.org/}%
\providecommand \selectlanguage [0]{\@gobble}%
\providecommand \bibinfo  [0]{\@secondoftwo}%
\providecommand \bibfield  [0]{\@secondoftwo}%
\providecommand \translation [1]{[#1]}%
\providecommand \BibitemOpen [0]{}%
\providecommand \bibitemStop [0]{}%
\providecommand \bibitemNoStop [0]{.\EOS\space}%
\providecommand \EOS [0]{\spacefactor3000\relax}%
\providecommand \BibitemShut  [1]{\csname bibitem#1\endcsname}%
\let\auto@bib@innerbib\@empty
%</preamble>
\bibitem [{\citenamefont {Rivas}\ and\ \citenamefont
  {Huelga}(2012)}]{rivas2012open}%
  \BibitemOpen
  \bibfield  {author} {\bibinfo {author} {\bibfnamefont {A.}~\bibnamefont
  {Rivas}}\ and\ \bibinfo {author} {\bibfnamefont {S.~F.}\ \bibnamefont
  {Huelga}},\ }\href@noop {} {\emph {\bibinfo {title} {Open quantum systems}}}\
  (\bibinfo  {publisher} {Springer},\ \bibinfo {address} {Berlin, Heidelberg},\
  \bibinfo {year} {2012})\BibitemShut {NoStop}%
\end{thebibliography}%
\end{document}

\title{Supplemantal Materials: Twisting Optomechanical Cavity}
\author{Daigo Oue}
\affiliation{Kavli Institute for Theoretical Sciences, University of Chinese Academy of Sciences, Beijing, 100190, China}
\affiliation{Department of Physics, Imperial College LondonPrince Consort Rd, Kensington, London SW7 2AZ, UK}
\author{Mamoru Matsuo}
\affiliation{Kavli Institute for Theoretical Sciences, University of Chinese Academy of Sciences, Beijing, 100190, China}
\affiliation{CAS Center for Excellence in Topological Quantum Computation, University of Chinese Academy of Sciences, Beijing 100190, China}
\date{\today}

\maketitle

In the Supplemental Materials,
we describe harmonic torsional oscillation, the Schwinger--Keldysh approach to calculate the rate of torsional oscillation pumping, single photon polarisation dynamics, and a possible experimental implementation.

\section{harmonic oscillator for the torsional oscillation}
For a rod with the mass density \(\rho\), the moment of inertia \(I\),
and the shear modulus \(S\), we consider torsional angle \(\theta\)
along the long direction. Let us write the equation of rotational motion
for a infinitesimal element \(\mathrm{d}x\) under a torque \(T\), 
\begin{align}
\rho I \mathrm{d}x \frac{\partial ^ 2}{\partial t ^ 2} \theta =
T + \frac{\partial T}{\partial x}\mathrm{d}x - T =
\frac{\partial T}{\partial x} \mathrm{d}x.
\end{align}
The torsional angle \(\theta\) and the torque \(T\) can be associated
with each other via the shear modulus \(S\), 
\begin{align}
T = SI\frac{\partial}{\partial x}\theta.
\end{align}
Substituting this relation into the equation of motion, we can obtain

\begin{align}
\bigg( 
\frac{\partial ^ 2}{\partial x ^ 2} -
\frac{1}{v ^ 2}\frac{\partial ^ 2}{\partial t ^ 2} 
\bigg) \theta = 0,
\end{align}
This is nothing but the wave equation with the sound velocity 
$v = \sqrt{\frac{S}{\rho}}$,
and thus, the torsional
oscillation may be modeled by a harmonic oscillator, i.e. 
\begin{align}
\frac{\partial ^ 2}{\partial t ^ 2} \theta_k =
-\omega_k ^ 2 \theta_k,
\quad 
\omega_k = vk.
\end{align}
As for the typical eigenfrequency of the torsional oscillation
\(\omega_k\), the order of magnitude can be up to \(100~\mathrm{GHz}\) in the case of
a quartz micro-rod ($S \sim 10 \ \mathrm{GPa}$, $\rho \sim 10 ^ 3\ \mathrm{kg/m ^ 3}$, $v \sim 100\ \mathrm{m/s}$, $k \sim 10\ \mathrm{\mu m ^ {-1}}$).

\section{Optically pumped torsional oscillation} 
Let us derive the pumping formula used in the text.
Our theory is based on the Schwinger--Keldysh approach.
After the rotating wave approximation, we have
\begin{align}
H_\mathrm{int} =
g(a_m J_+ + \mathrm{h.c.}),
\end{align}
where \(J_+ = a_e a_o^\dagger\) and $g$ is the interaction strength derived in the main text.
Note that this approximation is valid in our case because the frequency of the torsional interaction is turned far below other frequencies, e.g.~by applying weak anisotropy.
We are interested in the pumping rate, i.e.~the mechanical energy change per unit time,
\begin{align}
\frac{\partial}{\partial t}
\hbar \omega_m
a_m^\dagger a_m
= -i \hbar \omega_m 
g (a_m J_+- \mathrm{h.c.}),
\end{align}
where we have applied the Heisenberg equation of motion,
\(\partial O/\partial t = [O,H]/(i\hbar)\).

Here, we apply the Schwinger--Keldysh formalism to perturbatively evaluate the quantum statistical average of the pumping rate,
\begin{align}
I_m (t_1) =
-2\hbar \omega_m g 
\operatorname{Im}
\langle a_m(t_1^+) J_+(t_1^-) \rangle,
\end{align}
where the superscripts on the arguments,
$+$ and $-$,
denote the forward and backward branches on the Keldysh contour.
The statistical average is taken over the initial state,
\(\langle O\rangle = \mathrm{tr}[\rho_0 O S_C]\),
with the scattering
operator stemming from the torsional optomechanical term, 
\begin{align}
S_C = \mathcal{T} e^{\int_C{H_\mathrm{int}(t')}/{(i\hbar)}\mathrm{d}t'},
\end{align}
where the interaction Hamiltonian $H_\mathrm{int}$ is integrated on the Keldysh contour $C$,
and $\mathcal{T}$ is the time ordering operator on the contour.

After expanding the scattering operator up to the first order in the interaction strength \(g\),
which is smaller than any other characteristic frequencies in our system such as the effective optical frequency $\Delta$,
we apply the Bloch--de Dominicis theorem to get
\begin{align}
\langle a_m (t_1^+) J_+ (t_1^-) \rangle = 
-ig\int_C
\langle \mathcal{T}
a_m(t_1) a_m^\dagger(t_2)
\rangle_0
\langle \mathcal{T}
J_+(t_1) J_-(t_2)
\rangle_0\mathrm{d}t_2,
\end{align}
where we shall substitute
\(\langle O \rangle_0 = \mathrm{tr}[\rho_0 O]\).
Applying the Langreth rule,
we can get the real-time representation, 
\begin{align}
I_m (t_1) =
-2\hbar \omega_m \cdot (\hbar g)^2
\operatorname{Re} \int \big(
\chi_{12}^\mathfrak{R} G_{21}^< +
\chi_{12}^< G_{21}^\mathfrak{A}
\big) \mathrm{d}t_2,
\label{eq:I_m (real-time)}
\end{align}
where we have defined the nonequilibrium Green's functions for mechanical and optical systems, 
\begin{align}
i\hbar \chi_{12} = 
\langle \mathcal{T} a_m(t_1) a_m^\dagger(t_2) \rangle_0,
\quad
i\hbar G_{12} = 
\langle \mathcal{T} J_-(t_1) J_+(t_2) \rangle_0,
\end{align}
whose components read 
\begin{align}
i\hbar \chi_{12}^< = 
\langle a_m^\dagger(t_2) a_m(t_1) \rangle_0,
\quad
i\hbar \chi_{12}^> = 
\langle a_m(t_1) a_m^\dagger(t_2) \rangle_0,
\quad
i\hbar \chi_{12}^\mathfrak{R} = 
\theta(t_1 - t_2)
(i\hbar \chi_{12}^> - i\hbar \chi_{12}^<),
\\
i\hbar G_{12}^< = 
\langle J_+(t_2) J_-(t_1) \rangle_0,
\quad
i\hbar G_{12}^> = 
\langle J_-(t_1) J_+(t_2) \rangle_0,
\quad
i\hbar G_{12}^\mathfrak{A} = 
\theta(t_2 - t_1)
(i\hbar G_{12}^< - i\hbar G_{12}^>).
\end{align}

Defining the nonequilibrium distribution difference, 
\begin{align}
\delta f_{\omega} =
\frac{\chi_\omega^<}{2i\operatorname{Im}\chi_\omega^\mathfrak{R}}
-\frac{G_\omega^<}{2i\operatorname{Im}G_\omega^\mathfrak{R}},
\end{align}
makes Eq.~\eqref{eq:I_m (real-time)} simpler. 
At the steady state, e.g.
\(\chi_{12}^\mathfrak{R} \rightarrow \int \chi_{\omega}^\mathfrak{R} e^{-i\omega (t_1 - t_2)} \mathrm{d}\omega/(2\pi)\),
we have
\begin{align}
I_m^\mathrm{ss} =
4 \hbar \omega_m \cdot (\hbar g)^2
\int 
\operatorname{Im} \chi_{\omega}^\mathfrak{R}
\operatorname{Im} G_{\omega}^\mathfrak{R}
\delta f_{\omega}
\frac{\mathrm{d}\omega}{2\pi}.
\end{align}
Note that the distribution difference vanishes at thermal equilibrium,
where two subsystems are balanced.
If we drive the system out of the equilibrium,
for example, by applying an external field or a temperature gradient,
the difference becomes finite.

The retarded components and the nonequilibrium distribution difference can be given in the frequency domain,
\begin{align}
\chi_\omega^\mathfrak{R} =
\frac{1/\hbar}{\omega - \omega_m + i\Gamma},
\quad
G_\omega^\mathfrak{R} =
\frac{\langle a_o^\dagger a_o - a_e^\dagger a_e \rangle_0/\hbar}{\omega - \Delta + i\kappa},
\quad
\delta f_\omega =
\frac{\langle a_e^\dagger a_e \rangle_0}{\langle a_e^\dagger a_e - a_o^\dagger a_o\rangle_0}
2\pi \delta n_\mathrm{in} \delta(\omega - \Delta_d),
\end{align}
where \(\Delta = \omega_o - \omega_e\) and \(\Delta_d = \omega_d - \omega_e\) are the effective frequencies of the cavity mode and the external drive.
Those expressions are derived in the following section.
Therefore, we can explicitly write the pumping rate at the steady state,
\begin{align}
I_m^\mathrm{ss} =
4 \hbar \omega_m \cdot (\hbar g)^2
\operatorname{Im}
\frac{1/\hbar}{\Delta_d - \omega_m + i\Gamma}
\operatorname{Im}
\frac{\langle a_e^\dagger a_e \rangle_0/\hbar}{\Delta_d - \Delta + i\kappa}
\delta n_\mathrm{in},
\end{align}

\section{Nonequilibrium Green's functions}

The nonequilibrium Green's function for the optical subsystem is composed of contributions from the two modes,
\begin{align}
  i\hbar G_{12} 
  = \langle \mathcal{T} J_-(t_1) J_+(t_2) \rangle_0
  = \langle \mathcal{T} a_e^\dagger(t_1) a_o(t_1) a_e(t_2) a_o^\dagger(t_2) \rangle_0,
  = \langle \mathcal{T} 
  a_e^\dagger(t_1) a_e(t_2) 
  \rangle_0
  \langle \mathcal{T} 
  a_o(t_1) a_o^\dagger(t_2)
  \rangle_0,
\end{align} 
where we have applied the Bloch--de Dominicis theorem.
The lesser component is given by 
\begin{align}
i\hbar G_{12}^< =
\langle \mathcal{T} 
a_e^\dagger(t_1^+) a_e(t_2^-) 
\rangle_0
\langle \mathcal{T} 
a_o(t_1^+) a_o^\dagger(t_2^-)
\rangle_0
=: i\hbar G_{e;21}^> \cdot i\hbar G_{o;12}^<,
\end{align} 
while the greater component is 
\begin{align}
i\hbar G_{12}^> =
\langle \mathcal{T} 
a_e^\dagger(t_1^+) a_e(t_2^-) 
\rangle_0
\langle \mathcal{T} 
a_o(t_1^+) a_o^\dagger(t_2^-)
\rangle_0
=: i\hbar G_{e;21}^< \cdot i\hbar G_{o;12}^>.
\end{align}
Therefore, we can decompose the retarded component, 
\begin{align}
G_{12}^\mathfrak{R} =
\theta(t_1 - t_2) (G_{12}^> - G_{12}^<)
= i\hbar (
G_{e;21}^< \cdot G_{o;12}^> - 
G_{e;21}^> \cdot G_{o;12}^<).
\end{align}

We can explicitly write the lesser and greater components for each optical mode,
\begin{align}
G_{\sigma;12}^< =
\langle a_\sigma^\dagger(t_2) a_\sigma(t_1) \rangle_0 =
\langle a_\sigma^\dagger a_\sigma \rangle_0
e^{-i\omega_\sigma (t_1 - t_2)},
\quad
G_{\sigma;12}^> =
\langle a_\sigma(t_1) a_\sigma^\dagger(t_2) \rangle_0 =
\langle a_\sigma a_\sigma^\dagger \rangle_0
e^{-i\omega_\sigma (t_1 - t_2)},
\end{align}
and thus the retarded component in the time domain, 
\begin{align}
i\hbar G_{12}^\mathfrak{R} =
\theta(t_1 - t_2) \langle 
a_o^\dagger a_o - a_e^\dagger a_e 
\rangle_0
e^{-i\Delta (t_1 - t_2)},
\end{align}
where we have written the frequency difference \(\Delta = \omega_o - \omega_e\).
Applying the Fourier transformation, we can obtain
\begin{align}
G_{\omega}^\mathfrak{R} =
\frac{\langle 
a_o^\dagger a_o - a_e^\dagger a_e 
\rangle_0/\hbar
}{ \omega - \Delta + i\kappa/2}
= \langle a_o^\dagger a_o - a_e^\dagger a_e 
\rangle_0
G_{o;\omega + \omega_e}^\mathfrak{R},
\end{align}
where we have phenomenologically introduced the cavity damping constant \(\kappa\).

\section{Nonequilibrium distribution difference}

When an external drive is applied to the system,
Green's functions vary from the ones in the absence of the drive.
In this section, we shall evaluate the variations.
If the external drive modulates the ordinary mode,
the variation may be written as
\begin{align}
  i\hbar \delta G_{12}^<
  &= i\hbar G_{e;21}^> \cdot i\hbar \delta G_{o;12}^<
  = i\hbar \langle a_e^\dagger a_e \rangle_0
  \int 
  \delta G_{o;\omega}^< e^{-i (\omega - \omega_e)(t_1 - t_2)}
  \mathrm{d}\omega,
  \\
  &= i\hbar \langle a_e^\dagger a_e \rangle_0
  \int 
  \delta G_{o;\omega + \omega_e}^< e^{-i \omega (t_1 - t_2)}
  \mathrm{d}\omega.
\end{align}
The Fourier transformation gives
\begin{align}
\delta G_{\omega}^< = 
\langle a_e^\dagger a_e \rangle_0
\delta G_{o;\omega + \omega_e}^<
= \langle a_e^\dagger a_e \rangle_0
\delta f_{o;\omega + \omega_e}
\cdot 
2i\operatorname{Im}G_{o;\omega + \omega_e}^\mathfrak{R},
\end{align}
and therefore we can write
\begin{align}
\delta f_\omega = 
\frac{\delta G_{\omega}^<}{2i\operatorname{Im}G_{\omega}^\mathfrak{R}}
= \frac{
\langle a_e^\dagger a_e \rangle_0
}{
\langle a_o^\dagger a_o - a_e^\dagger a_e \rangle_0
}
\delta f_{o;\omega + \omega_e}.
\end{align}
Note that we have introduced the distribution difference for the ordinary mode,
$\delta f_{o;\omega} :=
\delta G_{o;\omega}^< / (2i\operatorname{Im}G_{o;\omega}^\mathfrak{R})$,
which is evaluated in the following.

Remind that the unperturbed part of our torsional optomechanical Hamiltonian is given by 
\begin{align}
H = H_\mathrm{em} + H_\mathrm{mech} =
\sum_{\sigma = e,o,m} \hbar \omega_\sigma a_\sigma^\dagger a_\sigma.
\end{align}
When an external drive is applied, we have an additional term accordingly.
If we drive the ordinary mode, we have a driving Hamiltonian in the following form:
\begin{align}
H_\mathrm{drv} = 
h(t) a_o(t) + h^*(t) a_o^\dagger(t),
\end{align}
where we defined the amplitude of the external drive
\(h(t) = -i \hbar \Omega_d e^{i\omega_d t}\)
with the driving strength
\(\Omega_d = \sqrt{P_\mathrm{in}\kappa/(\hbar \omega_d)}\).
Here, the driving power is denoted by \(P_\mathrm{in}\),
and the cavity damping constant \(\kappa\).
Remind that we can selectively modulate cavity modes by properly choosing the polarisation of the driving field because they are mutually orthogonal.

To find the nonequilibrium distribution function,
we evaluate the nonequilibrium Green's function for the ordinary mode in the presence of the driving term, 
\begin{align}
\langle \mathcal{T} a_o(t_1) a_o^\dagger (t_2) \rangle.
\end{align}
in which the statistical average is taken over the initial state with the scattering operator,
\(\langle O\rangle = \mathrm{tr}[\rho_0 O S_C]\).
The scattering operator representing the external drive is given by 
\begin{align}
S_C = \mathcal{T} e^{\int_C{H_\mathrm{drv}(t')}/{(i\hbar)}\mathrm{d}t'}. 
\end{align} 
Expanding the scattering operator,
we can evaluate the variation from the nonequilibrium Green's function in the absence of the external
drive.
Up to the second order in the driving strength,
we have
\begin{align}
\delta G_{o;12} :=
\langle \mathcal{T} a_o(t_1) a_o^\dagger (t_2) \rangle -
\langle \mathcal{T} a_o(t_1) a_o^\dagger (t_2) \rangle_0
\simeq \iint_C 
G_{o;13} \Sigma_{34} G_{o;42}
\mathrm{d}t_3 \mathrm{d}t_4,
\end{align}
where the Bloch-de Dominicis theorem has been applied so that only relevant pairs of operators remain in the equation. 
Note that we shall substitute \(\langle O \rangle_0 = \mathrm{tr}[\rho_0 O]\).
Note also that we have written the nonequilibrium Green's function in the absence of the external drive, 
\begin{align}
i\hbar G_{o;12} =
\langle 
\mathcal{T} a_o(t_1) a_o^\dagger(t_2)
\rangle_0,
\end{align}
and the contribution from the external drive, 
\begin{align}
i\hbar \Sigma_{34} =
\langle 
\mathcal{T} h(t_3) h^*(t_4)
\rangle_0.
\end{align}

Applying the Langreth rule, we can obtain the real-time representation,
\begin{align}
\delta G_{o;12}^< = \iint 
G_{o;13}^\mathfrak{R} \Sigma_{34}^< G_{o;42}^\mathfrak{A}
\mathrm{d}t_3 \mathrm{d}t_4.
\end{align}
In the frequency domain,
we have 
\begin{align}
\delta G_{o;\omega}^< =
G_{o;\omega}^\mathfrak{R} \Sigma_{\omega}^< G_{o;\omega}^\mathfrak{A}.
\end{align}
Note that the retarded component is given by 
\begin{align}
G_{o;\omega}^\mathfrak{R} =
\frac{1/\hbar}{\omega - \omega_o + i \kappa},
\end{align}
and the advanced component is obtained via
\([G_o^\mathfrak{A}(\omega)]^* = G_o^\mathfrak{R}(\omega)\).
Note also that we have 
\begin{align}
\Sigma_{\omega}^<  =
-2\pi i\hbar \Omega_d^2
\delta(\omega - \omega_d).
\end{align}

Therefore, the nonequilibrium distribution difference for the ordinary photons induced by the external drive is evaluated up to the leading order, 
\begin{align}
\delta f_{o;\omega} =
\frac{\delta G_{o;\omega}^< }{2i\operatorname{Im} G_{o;\omega}^\mathfrak{R}}
=\frac{\Sigma_{\omega}^</\hbar}{-i\kappa} =
2\pi \delta n_\mathrm{in} \delta(\omega - \omega_d)
\end{align}
Remind the delta function stemmed from integration over the time variable so that its unit is \(\mathrm{[s]}\),
and the distribution difference gives a dimensionless quantity after being integrated over the frequency as it should be.
The factor
\(\delta n_\mathrm{in} = P_\mathrm{in}/(\hbar \omega_d)\) 
in the equation corresponds to the number of photons pumped into the cavity per unit time.

\section{Single-photon polarisation dynamics}
For a single photon, 
two different polarisation states,
the extraordinary ray and ordinary ray modes, 
can be described by an effective two level system coupled to mechanical oscillation via the torsional optomechanical interaction.
Then, we can replace
\begin{align}
  \label{eq:replace_operators}
  \hat{a}_o^\dagger\hat{a}_e \mapsto \Sp,\quad  
  \hat{a}_o\hat{a}_e^\dagger \mapsto \Sm,\quad
  \sum_{\sigma=o,e}\hbar\omega_\sigma \hat{a}_\sigma^\dagger \hat{a}_\sigma \mapsto \frac{\Delta \omega}{2}\Sz,
\end{align}
where $\Sz$ is corresponding Pauli matrix,
and we have defined $\Delta \omega = \omega_o - \omega_e$ and $\Spm = \frac{(\Sx \pm i\Sy)}{2}$.
Note that we work in the unit $\hbar = 1$ in this section.
We get
\begin{align}
  \label{eq:H_tot_transformed}
  \hat{H} &= 
  \frac{\Delta\omega}{2}\Sz 
  + \omega_0 \hat{b}^\dagger \hat{b}
  + g(\Sp \hat{b} + \Sm \hat{b}^\dagger),
\end{align}
Here we have ignored the anti-resonance terms $g(\Sp\hat{b}^\dagger + \Sm\hat{b})$ with the help of the rotating wave approximation \cite{rivas2012open}.

\par
\begin{figure}[htbp]
  \centering
  \includegraphics[width=0.5\linewidth]{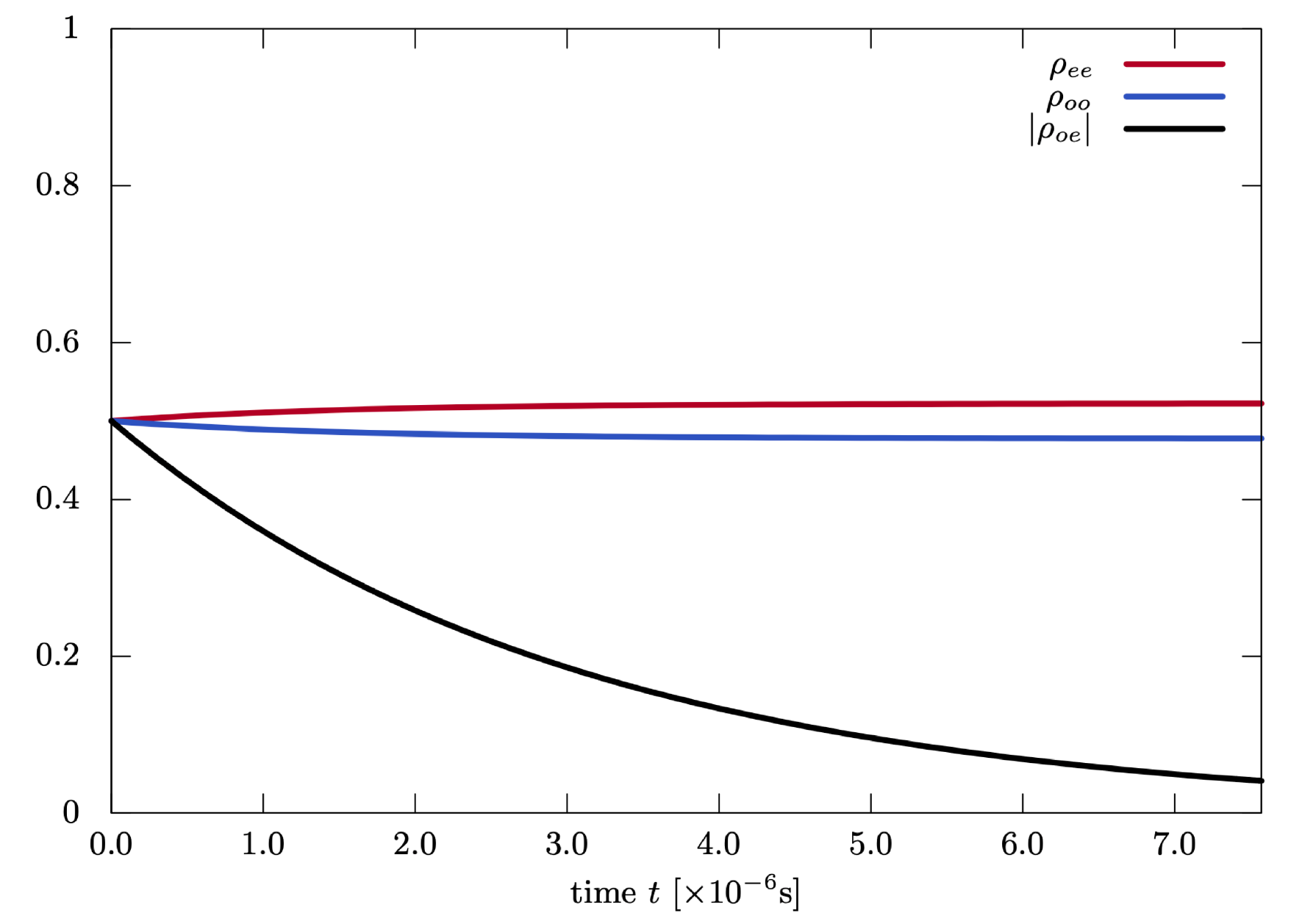}
  \caption{
    Population and coherence dynamics of the effective two level system.
    As an initial state,
    we prepare a diagonally polarised photon in the cavity \eqref{eq:rho_initial}.
    As time goes by, the offdiagonal elements of the density matrix decay to vanish,
    while the diagonal elements converge to finite values.
    Here we use refractive indices of quartz ($n_o=1.547,\ n_e=1.556$) and set $T = 300\ \mathrm{K}$ and cavity length $L=10^{-4}\ \mathrm{cm}$.
  }
  \label{fig:population_coherence}
\end{figure}

If we immerse the cavity in a gaseous environment (,
the torsional oscillation will be damped by the frictional drag and have a finite lifetime;
hence, the peak in the spectrum of the mechanical oscillation has a finite width.
In other words, the spectrum of the mechanical subsystem becomes continuous rather than discrete.
In this case,
we consider not only a single mode but many modes of mechanical oscillation couple to the photonic modes in the cavity.
We label each mode by its frequency $\omega$.
The Hamiltonian becomes
\begin{align}
  \label{eq:H}
  \hat{H} &= \frac{\Delta\omega}{2}\Sz 
  + \int_0^{\omega_{\mathrm{max}}} \hat{b}_\omega^\dagger \hat{b}_\omega \mathrm{d}\omega
  + \int_0^{\omega_{\mathrm{max}}} g(\omega)(\Sp \hat{b}_\omega + \Sm \hat{b}_\omega^\dagger)
  \mathrm{d}\omega,
\end{align}
where the coupling $g(\omega)$ can be obtained by following the same derivation described in the main text for each torsional oscillation $\hat{b} _ \omega$.
By tracing out the mechanical oscillation degree of freedom in the weak coupling limit,
we obtain the Lindblad master equation \cite{rivas2012open},
\begin{align}
  \notag
  \frac{d \hat{\rho}(t)}{dt} 
  &= -i \left[
    \left( \frac{\Delta \omega}{2} + \delta\right) \Sz, \hat{\rho}(t)
  \right]\\ \notag
  &+ \left(n(\Delta \omega)+1\right) \gamma
  \left[
    \Sm \hat{\rho}(t) \Sp - \frac{1}{2}\left\{ \Sp \Sm, \hat{\rho}(t)\right\}
  \right] \\ \label{eq:master_eq}
  &+ n(\Delta \omega) \gamma 
  \left[
    \Sp \hat{\rho}(t) \Sm - \frac{1}{2}\left\{ \Sm \Sp, \hat{\rho}(t)\right\}
  \right],
\end{align}
which describes the photon polarisation dynamics in the twisted optical cavity.
Here, $n(\Delta \omega) = 1/(e^{\beta \Delta \omega} - 1)$ is the average number of bosons at the frequency $\Delta \omega$ and the inverse temperature $\beta$,
$\gamma = 2\pi \left\{g(\Delta \omega)\right\}^2$ is a decay coefficient,
and $\delta$ is a frequency shift induced by thermal and zero-point fluctuations of the mechanical modes,
\begin{align}
  \label{eq:delta}
  \delta
  &= 
  \mathcal{P}\int_0^{\omega_{\mathrm{max}}} 
  \frac{\left\{g(\omega)\right\}^2}{\Delta \omega - \omega}
  \left(n(\omega) + \frac{1}{2}\right)
  \mathrm{d}\omega,
\end{align}
where $\mathcal{P}$ means the principal value of the integration.
The density operator is a $2 \times 2$ matrix whose diagonal elements correspond to the photon population in each mode and offdiagonal elements represent the coherence between the two modes.
In \figref{fig:population_coherence},
we show the photon polarisation dynamics in the twisted optomechanical cavity.
We prepare a photon in the diagonal polarisation state inside the cavity,
which is a coherent superposition of the ordinary ray and the extraordinary ray modes.
The corresponding initial value of the density matrix is
\begin{align}
  \label{eq:rho_initial}
  \hat{\rho}(0) = 
  \begin{pmatrix}
    1/2 & 1/2\\
    1/2 & 1/2
  \end{pmatrix}.
\end{align}
According to the master equation,
the diagonal elements of the density matrix converge to steady state values,
\begin{align}
  \label{eq:steady}
  \rho_{oo} \rightarrow \frac{n(\Delta \omega)}{2n(\Delta \omega)+1},\quad
  \rho_{ee} \rightarrow \frac{n(\Delta \omega)+1}{2n(\Delta \omega)+1},
\end{align}
while the offdiagonal elements decay towards zero.
From these observations, 
the photon experiences the decoherence (depolarisation) and ends up with an incoherently mixed state.

\section{Possible experimental implementation} 
\begin{figure}[htbp]
  \centering
  \includegraphics[width=.5\linewidth]{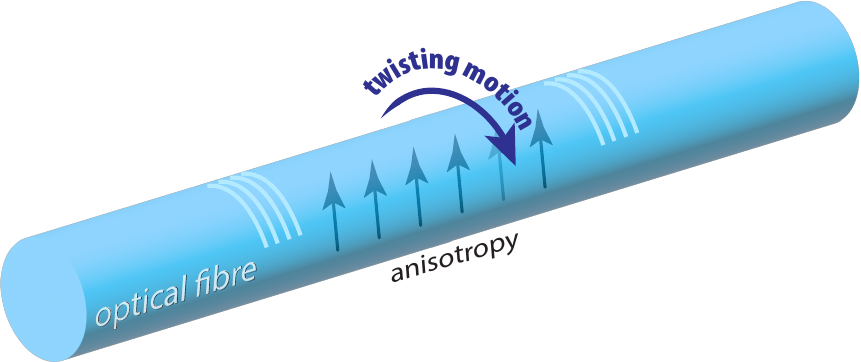}
  \caption{
    A possible implementation of the proposed torsional optomechanics.
    One can fabricate reflectors on an optical fibre with a small separation in between,
    where electromagnetic radiation can be confined.
    If the fibre has anisotropy, the confined optical modes induce twisting motion.
    The desired anisotropy can be played not only by the optical property of the medium but also effectively by the geometry, e.g.~elliptical core.
  }
  \label{fig:fibre}
\end{figure}

Instead of preparing a cavity and inserting an anisotropic rod, one could alternatively make a cavity on an anisotropic fibre as shown in \figref{fig:fibre}.
If one adopts such a fibre setup, the desired anisotropy can be played not only by the optical property of the medium but also effectively by the geometry, e.g.~elliptical core. 

\bibliography{twist}